\documentclass[twocolumn,showkeys,preprintnumbers,amsmath,amssymb]{revtex4}
	\usepackage[latin1]{inputenc}
	\usepackage{graphicx}
	\usepackage{listings}
	\usepackage{lipsum}

\begin{document}

	\title{Evolutionary aspects of Reservoir Computing}

	\providecommand{\CSL}{ICREA-Complex Systems Lab, Universitat Pompeu Fabra, 08003 Barcelona}
	\providecommand{\IBE}{Institut de Biologia Evolutiva (CSIC-UPF), 08003 Barcelona. }
	\author{Lu\'is F Seoane} \affiliation{\CSL} \affiliation{\IBE}

	\vspace{0.4 cm}

	\begin{abstract}
		\vspace{0.2 cm}

		Reservoir Computing (RC) is a powerful computational paradigm that allows high versatility with cheap learning.
		While other artificial intelligence approaches need exhaustive resources to specify their inner workings, RC is
		based on a reservoir with highly non-linear dynamics that does not require a fine tuning of its parts. These
		dynamics project input signals into high-dimensional spaces, where training linear readouts to extract input
		features is vastly simplified. Thus, inexpensive learning provides very powerful tools for decision making,
		controlling dynamical systems, classification, etc. RC also facilitates solving multiple tasks in parallel,
		resulting in a high throughput. Existing literature focuses on applications in artificial intelligence and
		neuroscience. We review this literature from an evolutionary perspective. RC's versatility make it a great candidate
		to solve outstanding problems in biology, which raises relevant questions: Is RC as abundant in Nature as its
		advantages should imply? Has it evolved? Once evolved, can it be easily sustained? Under what circumstances? (In
		other words, is RC an evolutionarily stable computing paradigm?) To tackle these issues we introduce a conceptual
		morphospace that would map computational selective pressures that could select for or against RC and other computing
		paradigms. This guides a speculative discussion about the questions above and allows us to propose a solid research
		line that brings together computation and evolution with RC as a working bench.

	\end{abstract}


	\maketitle

	\section{Introduction}
		\label{sec:1}

		Somewhere between pre-biotic chemistry and the first complex replicators, information assumed a paramount role in
		our planet's fate \cite{SzathmaryMaynardSmith1997, Joyce2002, WalkerDavies2013}. From then onwards, Darwinian
		evolution explored multiple ways to organize the information flows that shape the biosphere \cite{Schuster1996,
		Smith2000, JablonkaLamb2006, Nurse2008, Joyce2012, Adami2012, HidalgoMaritan2014, SmithMorowitz2016}. As Hopfield
		argues, ``biology looks so different'' because it is ``physics plus information'' \cite{Hopfield1994}. Central in
		this view is the ability of living systems to capitalize on available external information and forecast regularities
		from their environment \cite{Jacob1998, Wagensberg2000}, a driving force behind life's progression towards more
		complex computing capabilities \cite{SeoaneSole2018a}.

\begin{figure*}
			\begin{center}
				\includegraphics[width = 0.75 \textwidth]{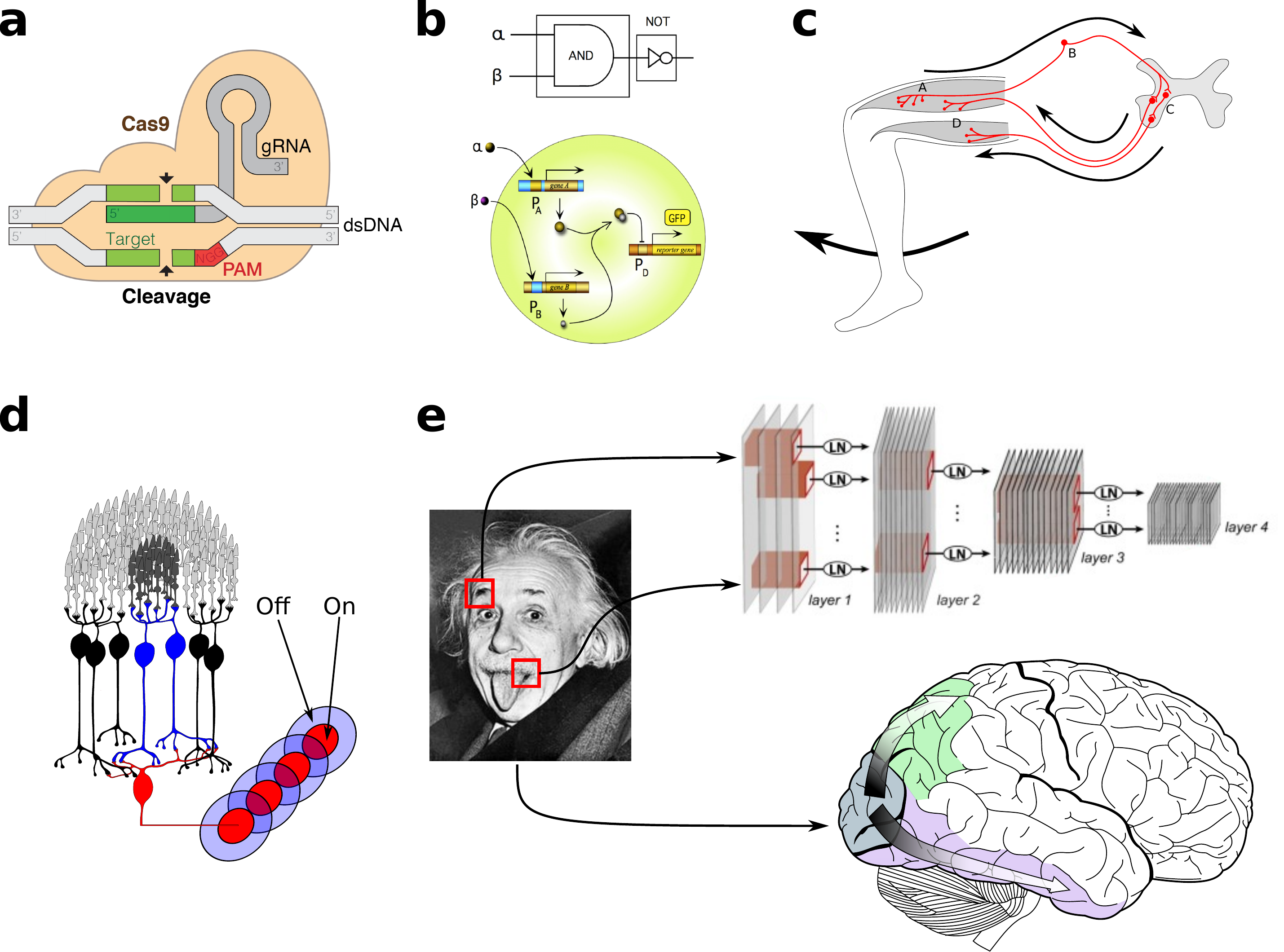}

        \caption{ {\bf Some computing devices in biology. } {\bf a} DNA and RNA performs multiple pattern recognition,
starting with the simplest matching of codons in ribosomes to synthesize proteins. More complex pattern matching (here,
Cas9 matches a longer string of RNA to DNA, drawing from Marius Walter available at
https://commons.wikimedia.org/w/index.php?curid=62766587) allows cleavage and insertion of DNA snippets. {\bf b} Boolean
logic determines gene expression or silencing. In the example, a receptor gene that expresses GFP is only active if
genes A and B are expressed and their corresponding products dimerize. This implements a Boolean AND gate. (Redrawn from
\cite{MaciaSole2014}.) {\bf c} The {\em knee jerk} reflex is implemented by sequential circuits that remind us of the
simple wiring of an electrical installation. Sensory neurons (A) gather a sudden input. Information follows the arrows
towards ganglia at the entrance (B) and gray matter of the spinal cord. There, information splits in two: it is
partially sent upstream towards the brain stem (not shown) and partially used to excite motor neurons (C) which right
away pull from muscles at their end (D). {\bf d} Retinal ganglion cells (red) pool information from rods and cones
(gray) through horizontal, bipolar (black and blue), and amacrine cells. This pooling involves excitatory (blue) and
inhibitory (black) mediating connections, which make a ganglion cell responsive to light and its absence in so-called
on- and off- centers and surround (red and blue discs, corresponding to dark and light gray cones). Receptive fields of
several ganglia are consequently pooled as well, resulting in edge detectors. {\bf e} Specific neurons implementing
average and pooling exist throughout the visual cortices \cite{FukushimaMiyake1982}. These are the building blocks of
deep convolutional neural networks \cite{KrizhevskyHinton2012}. The overall information flow, architecture, and various
receptive fields of such artificial networks reminds several processing steps found in mammal visual cortices
\cite{YaminsDiCarlo2014}, including edge detection in V1 thanks to retinal ganglia. (This image was drawn with elements
from \cite{YaminsDiCarlo2014} and https://commons.wikimedia.org/w/index.php?curid=1679336.) }

				\label{fig:1}
			\end{center}
		\end{figure*}

		We can trace computation in biology from pattern recognition in RNA and DNA \cite{PaunSalomaa2005, Doudna2017}
		(figure \ref{fig:1}{\bf a}), through the Boolean logic implemented by interactions in Gene Regulatory Networks
		\cite{Thomas1973, Kauffman1996, RodriguezCasoSole2009} (figure \ref{fig:1}{\bf a}), to the diverse and versatile
		circuitry implemented by nervous systems of increasing complexity \cite{DayanAbbott2001, Seung2012} (figure
		\ref{fig:1}{\bf c-e}). Computer Science, often inspired by biology, has reinvented some of these computing
		paradigms; usually from simplest to most complex, or guided by their saliency in natural systems. It is no surprise
		that we find some fine-tuned, sequential circuits for motor control (figure \ref{fig:1}{\bf c}) that resemble the
		wiring of electrical installations. Such pipelined circuitry gets assembled to perform parallel and more 
		coarse-grained operations, e.g., in assemblies of ganglion retinal cells that implement {\em edge detection}
		\cite{Levick1967, RussellWerblin2010} (figure \ref{fig:1}{\bf d}) similarly to filters used in image processing
		\cite{MarrHildreth1980, Marr1982, StephensBialek2013}. Systems at large often present familiar design philosophies
		or overall architectures, as illustrated by the resemblance between much of our visual cortex (figure
		\ref{fig:1}{\bf e}) and deep convolutional neural networks for computer vision \cite{FukushimaMiyake1982,
		KrizhevskyHinton2012, YaminsDiCarlo2014, KhalighRazaviKriegeskorte2014} (figure \ref{fig:1}{\bf f}).

		Such convergences suggest that chosen computational strategies might be partly dictated by universal pressures. We
		expect that specific computational tricks are readily available for natural selection to exploit them (e.g.
		convolving signals with a filter is faster in Fourier space, and the visual system could take advantage of it). Such
		universalities could constrain network structure in specific ways. We also expect that the substrate chosen for
		implementing those computations is guided by what is needed and available. This is, at large, one of the topics
		discussed in this volume. Different authors explore specific properties of computation as implemented, on the one
		hand, by {\em liquid} substrates with moving components such as ants or T-cells; and, on the other hand, by {\em
		solid brains} such as cortical or integrated circuits. Rather than this `thermodynamic state' of the hardware
		substrate, this paper reviews the Reservoir Computing (RC) framework \cite{Jaeger2001, MaassMarkram2002,
		JaegerPrincipe2007, VerstraetenStroobandt2007, LukoseviciousSchrauwen2012}, which somehow deals with a `solid' or
		`liquid' quality of the signals involved; hence rather focusing on the `state' of the software. As with other
		computing architectures, tricks, and paradigms, we expect that the use of RC by Nature responds to evolutionary
		pressures and contingent availability of resources.

		RC matters within a broader historical context because it has helped bypass a huge problem in machine learning. The
		first widely successful artificial intelligence architecture were layered, feed-forward networks (figure
		\ref{fig:1}{\bf f} and \cite{KrizhevskyHinton2012} are modern examples). These get a static input (e.g. a picture)
		whose basic features (intensity of light in each pixel) are read and combined by artificial neurons or units. A
		neuron's reaction, or activation, is determined by a collection of weights that measure how much each of the
		features matters to that unit. These activations are conveyed forward to newer neurons that use their own weights to
		combine features non-linearly, and thus extract complex structures (edges, shapes, faces, ...). Eventually, the
		whole network settles into a fixed state. A set of output units returns the static result of a computation (e.g.
		whether Einstein is present in the picture). Training a network consists in adjusting the weights for every neuron
		such that the system at large implements a desired computation (e.g. automatic face classification). Solving this
		problem in feed-forward networks with several layers was a challenge for decades until a widespread solution 
		(back-propagation \cite{RumelhartWilliams1986}) was adopted. Recurrent Neural Networks (RNN) brought in more
		computational power to the field. RNN contemplate feedback from more forward to earlier processing neurons. These
		networks do not necessarily settle in static output states, allowing them to produce dynamic patterns, e.g. for
		system control. They are also apt to process spatiotemporal inputs (videos, voice recordings, temporal data series,
		...), finding dynamical patterns often with long-term dependencies. Echoing the early challenges in feed- forward
		networks, full RNN training (i.e. adjusting every weight optimally for a desired task) presents important problems
		still not fully tamed \cite{BengioFrasconi1994, PascanuBengio2018}.

		RC is an approach that vastly simplifies the training of RNN, thus making more viable the application of this
		powerful technology. Instead of attempting to adjust every weight in the network, RC considers a fixed {\em
		reservoir} that does not need training (figure \ref{fig:2}{\bf a}) which works as if multiple, parallel
		spatiotemporal filters were simultaneously applied onto the input signal. This effectively projects non-linear input
		features onto a huge-dimensional space. There, separating these features becomes a simple, linear task. Despite the
		simplicity of this method, RC-trained RNN have been robustly used for a plethora of tasks including data
		classification \cite{VerstraetenStroobandt2006, JaegerSiewert2007, SoriaRuffini2018}, systems control
		\cite{JoshiMaass2004, SalmenPloger2005, Burgsteiner2005, JaegerSiewert2007}, time-series prediction
		\cite{JaegerHaas2004, IbanezSoriaRuffini2018}, uncovering grammar and other linguistic and speech features
		\cite{VerstraetenVanCampenhout2005, JaegerSiewert2007, TongCottrell2007, TriefenbachMartens2010, HinautDominey2013}
		etc.

		Again, we expect that Nature has taken advantage of any computing approaches available, including RC, and that
		important design choices are affected by evolutionary constraints. These will be major topics through the paper: how
		could RC be exapted by living systems and how might evolutionary forces have shaped its implementation. Section
		\ref{sec:2} provides a brief introduction to RC and reviews what its operating principles (notably optimal reservoir
		design) imply for biological systems. Section \ref{sec:3} shows inspiring examples from biology and engineering.
		Comments on selective forces abound across the paper, but section \ref{sec:4} wraps up the most important messages.
		This last part is largely speculative in an attempt to pose relevant research questions and strategies around RC,
		evolution, and computation. We will hypothesize about two important topics: i) what explicit evolutionary conditions
		might RC demand and ii) what evolutionary paths can transform a system into a reservoir.

	\section{Computational aspects of Reservoir Computing} 
		\label{sec:2}

		\subsection{Reservoir Computing in a nutshell} 
			\label{sec:2.1} 

			RC was simultaneously introduced by Herbert Jaeger \cite{Jaeger2001} and Wolfgang Maass and his colleagues
			\cite{MaassMarkram2002}. Jaeger arrived to {\em Echo State Networks} from a machine learning approach while Maass
			et al. developed {\em Liquid State Machines} with neuroscientifically realistic spiking neurons. The powerful
			operating principle is the same behind both approaches, later unified under the RC label \cite{JaegerPrincipe2007,
			VerstraetenStroobandt2007, LukoseviciousSchrauwen2012}.

			Consider a Recurrent Neural Network (RNN) consisting of $N$ units, all connected to each other, which receive an
			external input $\bar{y}^{in}(t) \equiv \{y^{in}_i(t), \> i=1, \dots, N\}$ (with $y^{in}_i(t) = 0$ if the $i$-th
			unit receives no input). Each unit has an internal state $x_i(t)$ that evolves following:   
				\begin{eqnarray}
					x_i(t+\Delta t) &=& \sigma \left( \sum_{j=1}^N \omega_{ij}x_j(t) + y^{in}_i(t) \right),      
					\label{eq:1}
				\end{eqnarray} 
			where $\sigma(\cdot)$ represents some non-linear function (e.g. an hyperbolic tangent). Variations of this basic
			theme appear in the literature. For example, continuous dynamics based on differential equations could be used; or
			inputs could consist of weighted linear combinations of $N^{feat}$ more fundamental features $\{u_k(t), k=1,
			\dots, N^{feat}\}$ such that $y^{in}_i(t) = \sum_{k=1}^{N^{feat}} \omega^{in}_{ik} u_k(t)$. This would allow us to
			trade off importance of external stimuli versus internal dynamics.

			Such RNN can be trained so that $N^{out}$ designated output units produce a desired response $\bar{y}^{out}(t)
			\equiv \{y_i^{out}(t), \> i=1, \dots, N^{out}\}$ when $\bar{y}^{in}(t)$ is fed into the network. The training
			consists in varying the $\omega_{ij}$ until the state of the output units given the input $\bar{x}^{out}(t |
			\bar{y}^{in}(t)) \equiv \{x^{out}_i(t | \bar{y}^{in}(t)), \> i=1, \dots, N^{out}\}$ matches the desired behavior
			$\bar{y}^{out}(t)$. A naive approach is to initialize the $\omega_{ij}$ randomly and modify all weights, e.g.
			using gradient descent, to minimize some error function $\epsilon(\bar{y}^{out}(t), \bar{x}^{out}(t |
			\bar{y}^{in}(t)))$ that measures how much does the network activity deviate from the target. Such training
			procedure is often useless because the RNN's recurrent dynamics introduce insurmountable numerical problems
			\cite{BengioFrasconi1994, PascanuBengio2018}.

		\begin{figure*}
			\begin{center}
				\includegraphics[width = 0.75\textwidth]{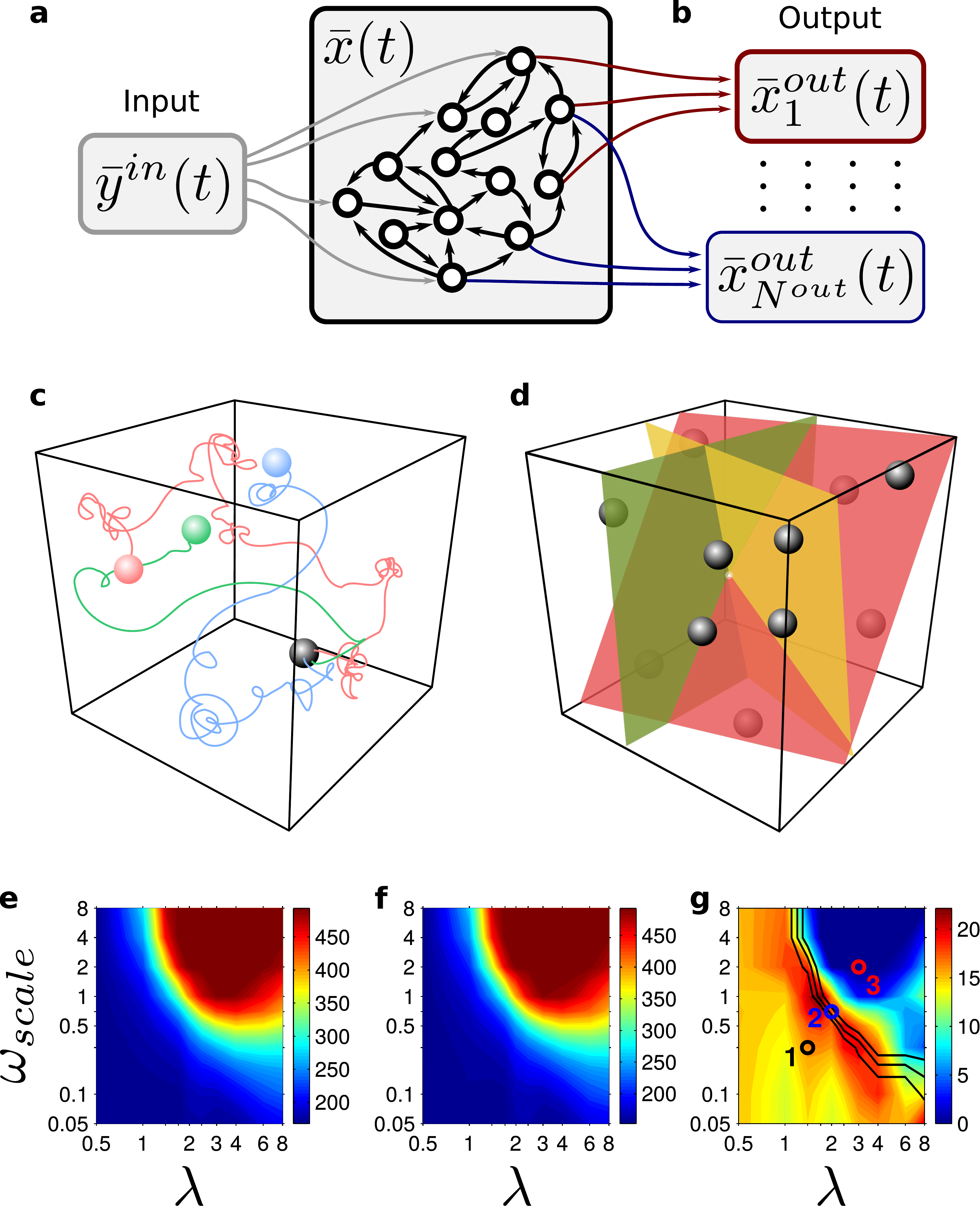}

        \caption{ {\bf Computational basis of reservoir computing. } {\bf a} Reservoir dynamics (black nodes and arrows)
are driven by external inputs (gray arrows). For RC, weights within the reservoir are not modified. {\bf b} A set of
$N^{out}$ output nodes can be appended. Each of them computes a linear combination of the reservoir state and solves a
single task. In RC training, only the weights of these output units are modified. This allows us to solve several (in
this case $N^{out}$) different tasks simultaneously without interference. {\bf c} The key to RC computing is that the
reservoir filters, and thus projects, low-dimensional input signals into a huge-dimensional space. Here we observe three
example trajectories (elicited, e.g., by three different inputs) through a representation of this abstract space. {\bf
d} Appropriate reservoir dynamics will make it possible to separate the projection of different inputs using simple
linear classifiers (implemented by the readout units from {\bf b}). If we sort in a matrix the different dynamical
states reached after different inputs, the rank of this matrix tells us how linearly independent the dynamics have
remained. This gives us an idea of how many different hyperplanes (i.e., readouts and different tasks) we can draw in
this abstract space. {\bf e-g} $r^S$, $r^G$, and $r^S-r^G$. Plot reproduced from \cite{LegensteinMaass2007b}. {\bf e}
$r^S$, which captures this separability property, is plotted for reservoirs made of simulated cortical columns with
varying correlation length between neurons ($\lambda$) and scale of synaptic strength ($\omega_{scale}$). We find large
$r^S$ (hence large separability) for chaotic dynamics achieved by large $\lambda$ and $\omega_{scale}$. {\bf f} When
noise and redundant information has been provided, we expect the rank of a similar matrix ($r^G$, which measures a
generalization capability) to score low. The same microcolumns as before present good generalization property when the
dynamics are ordered and similar inputs are consistently mapped into a same region of the space of reservoir dynamics.
{\bf g} Subtracting $r^G$ from $r^S$ measures a balance between the two desired reservoir properties. Good performance
also correlates with a measure of criticality (dot marked `2', where the Lyapunov exponent vanishes). Performance is bad
for too ordered (dot `1', negative Lyapunov exponent) or too chaotic dynamics (dot `3', positive Lyapunov exponent).
Black curves on top indicate the limits where Lyapunov exponents are fairly close to $0$, thus marking the edge of
chaos. }

				\label{fig:2}
			\end{center}
		\end{figure*}

			The RC approach to RNN training still uses random $\omega_{ij}$ but does not attempt to modify them. We say then
			that the units described by equation \ref{eq:1} constitute a {\em reservoir} (figure \ref{fig:2}{\bf a}). Upon it,
			we append a set of $N^{out}$ {\em readout units} (figure \ref{fig:2}{\bf b}) whose activity $\bar{x}^{out}(t)
			\equiv \{x^{out}_i(t), \> i=1, \dots, N^{out}\}$ is just a linear combination of the reservoir activity:
				\begin{eqnarray}
					x^{out}_i(t) &=& \sum_{j=1}^{N} \omega^{out}_{ij}x_j(t). 
					\label{eq:2}
				\end{eqnarray}

			Training proceeds on these output units alone. Only the $\omega^{out}_{ij}$ are modified; these do not feed back
			into the reservoir, which remains unchanged. The absence of this feedback during learning dissolves the grave
			numerical problems that affect other RNN. Finding the right $\omega^{out}_{ij}$ for a task becomes as simple as a
			linear regression between reservoir activity over time given an input $x_j(t | \bar{y}^{in}(t))$ and the desired
			target $\bar{y}^{out}(t)$ \cite{WyffelsStroobandt2008}. For a review on RC training (including issues of reservoir
			design discussed in the next section) see \cite{LukoseviciousJaeger2009} and other works \cite{Jaeger2002,
			Lukosevicious2012}. Also, the computational capacity of RC can be boosted if a (trainable) feedback is added from
			the linear readouts to the reservoir \cite{MaassSontag2007, SussilloAbbott2009, DaiHarley2009, RivkindBarak2017,
			CeniLivi2018}. This allows for more agile context-dependent computations and longer-term memory
			\cite{MaassSontag2007}. Controlling the network attractors (thus the learning) that such feedback induces is still
			not fully understood. These and other variations upon the RC theme are good steps towards full RNN training.

			Back to the most naive version of RC, the only task of the reservoir is to have its dynamic internal state
			perturbed by the input. In doing so, through its non-linear, convoluted dynamics, the reservoir is picking up the
			external signal and projecting it into the huge-dimensional space that consists of all possible dynamic
			configurations of the reservoir (figure \ref{fig:2}{\bf c}). This  high-dimensional space hopefully renders
			relevant features from the input more easily separable. Ideally, such features could be separated with a simple
			hyperplane that bisects this abstract space. This is precisely what the $\omega^{out}_{ij}$ implement. RC training
			consists in finding the right hyperplane solving each task given the reservoir.

			Of course, very poor dynamics could project inputs into boring reservoir configurations so that prominent features
			cannot be picked up. Next section discusses issues of reservoir design so that it produces optimal dynamics.
			However, the most important part of the training falls upon the readouts, whose dynamics do not affect the
			reservoir. This brings in the two most important advantages of RC: i) Learning is extremely easy, as just
			discussed, and ii) multiple readouts can be appended to a same reservoir (thus solving different, parallel tasks)
			without interference.

		\subsection{Good reservoir design: separability, generalization, criticality, and chaos}
			\label{sec:2.2} 

			When RC was first introduced, it was requested that the reservoir dynamics fulfilled a series of theoretical
			conditions such as presenting echo states, fading memory, separation property, etc \cite{Jaeger2001,
			MaassMarkram2002}. These are formal requirements to prove relevant computational theorems. In practice, a wide
			variety of systems can trivially work as reservoirs. As proofs of concept, among many others, in silico
			implementations have used realistic theoretical models of neural networks \cite{MaassMarkram2002,
			MaassBertschinger2005, LegensteinMaass2007b}, networks of springs \cite{HauserMaass2011, NakajimaPfeifer2013a}, or
			cellular automata \cite{NicheleGundersen2017}; and in hardware, analog circuits \cite{SorianoVanDerSande2015,
			DuLu2017}, a bucket of water \cite{FernandoSojakka2003}, and diverse photonic devices \cite{AppletantFischer2011,
			PaquotMassar2012, VandoorneBienstman2014} have been tried. All those theoretical conditions on reservoir dynamics
			boil down to two desired properties: i) reservoir dynamics must be able to separate different input features that
			are meaningful for a variety of tasks while ii) these same dynamics must be able to generalize to unseen examples,
			thus projecting reasonably similar inputs into a reasonable neighborhood within the dynamical space of the
			reservoir. To fulfill these conditions we want our reservoirs to behave somehow in between chaotic and simple
			dynamics, thus resonating with ideas about criticality in complex systems, as discussed below. \\

			In a series of papers \cite{MaassBertschinger2005, LegensteinMaass2007b, LegensteinMaass2007a}, Maass et al.
			explored two elegant measures that quantify the ability of a reservoir to {\em separate} relevant features and to
			{\em generalize} to unseen examples.

			Regarding separability, since RC works with simple linear readouts, we can quantify straightforwardly how many
			different binary features can be extracted by hyperplanes that bisect the space of dynamic configurations of the
			reservoir (figure \ref{fig:2}{\bf d}). Suppose a collection of inputs $Y^{S} \equiv \{y^{S}_k(t),\>k=1, \dots,
			N^{S}\}$ is fed to the reservoir. At time $t_0 + T$ after input onset (which happened at $t_0$), the reservoir
			activity $\bar{x}(t_0+T | y^{S}_k) \equiv \{x_i(t_0+T | y^{S}_k), \> i=1, \dots, N\}$ is recorded in an $N$-sized
			array. The collection $X^{S}(t_0+T | Y^{S}) \equiv \{\bar{x}(t_0+T | y^{S}_k), \> k=1, \dots, N^{S}\}$ consists of
			$N^{S}$ such arrays sorted in a matrix. The rank $r^S \equiv rank\{X^{S}\}$ conveys an idea of how linearly
			independent the driven activity of the reservoir is -- i.e. of how many  binary-classifiable features the
			reservoir can pick apart. Given $Y^{S}$, there are $2^{N^{S}}$ different possible binary classification problems.
			If $r^S = N^{S}$ we ensure that the dynamics of the reservoir can pick apart the features relevant to all these
			problems. Even if $r^S < N^{S}$, in general, the larger $r^S$ the better a reservoir would be, since it would make
			more degrees of freedom available to set up the linear readouts.

			As for a reservoir's ability to generalize, we approach the problem similarly, but assuming now that our input
			contains some redundant or tangential information (e.g. some of its variability comes from noise). Good reservoirs
			should be able to smooth this out. Assume that a larger collection of inputs $Y^{G} \equiv \{y^{G}_k(t),\>k=1,
			\dots, N^{G} > N^{S}\}$ is fed to the reservoir. Its dynamics are similarly captured by the matrix $X^{G}(t_0+T |
			Y^{G}) \equiv \{\bar{x}(t_0+T | y^{G}_k), \> k=1, \dots, N^{G}\}$. If the reservoir is capable of classifying
			noisy versions of the input under a same class (i.e. of generalizing), we expect now the rank $r^G \equiv
			rank\{X^{G}\}$ to be as small as possible. (More rigorous analysis relate $r^G$ to the $VC$-dimension of the
			system. This is, to the volume of input space that can be {\em shattered} by the reservoir dynamics -- see
			\cite{MaassBertschinger2005, LegensteinMaass2007b, LegensteinMaass2007a, Vapnik1998, CherkasskyMulier1998} for
			further explanations.)

			These measures rely on the expectation that the rank of arbitrary activity matrices (such as $X^S$ and $X^G$) will
			increase if new meaningful examples are provided, but will stall if examples add spurious or redundant
			variability. It is an open question, which still depends on the eventual task, what constitutes meaning and noise
			in each case. We expect to find more spurious information the more examples we have -- hence $N^G > N^S$. Final
			values of $r^S$ and $r^G$ will still depend on $N^S$ and $N^G$ -- which could stand, e.g., for test and train set
			sizes.

			Despite these caveats, naive applications of $r^S$ and $r^G$ seem to capture good reservoir design that works for
			a variety of problems. In \cite{MaassBertschinger2005, LegensteinMaass2007b, LegensteinMaass2007a}, realistic
			models of cortical columns were used as reservoirs. These models mimic the $3$-D geometric disposition of neurons
			in the neocortex. Given two neurons located at $\overrightarrow{x}_1$ and $\overrightarrow{x}_2$, the likelihood
			that they are connected decays as $\exp\left(-D^2(\overrightarrow{x}_1, \overrightarrow{x}_2)/\lambda^2\right)$
			with the Euclidean distance $D(\overrightarrow{x}_1, \overrightarrow{x}_2)$ between them. The parameter $\lambda$
			introduces an average geometric length of connections, resulting in sparse or dense circuits if $\lambda$ is
			respectively small or large. This in turn leads to short lived or more sustained dynamics. Individual neurons were
			simulated with equations for realistic leaky membranes \cite{MarkramTsodyks1998}, including proportions of
			inhibitory neurons within the circuit. Synaptic strengths were drawn randomly to reflect biological data (see
			\cite{MaassMarkram2002}), and they all were scaled by a common factor $\omega_{scale}$. Small $\omega_{scale}$ led
			to weakly coupled neurons in which activity faded quickly; while large $\omega_{scale}$ implied strong coupling
			between units, resulting in more active dynamics.

			The authors produce morphologically diverse reservoirs by varying $\lambda$ and $\omega_{scale}$. Too sparse a
			connection (due to low likelihood of connection or weak synapses -- respectively low $\lambda$ and
			$\omega_{scale}$, lower-left corner in figures \ref{fig:2}{\bf e-g}) leads to poor dynamics. This results in
			undesired low separability (low $r^S$, figure \ref{fig:2}{\bf e}) but brings about the expected large
			generalization (small $r^G$, figure \ref{fig:2}{\bf f}) because large classes of noisy input result in converging
			reservoir dynamics. Meanwhile strong synapses and a very dense network (large $\lambda$ and $\omega_{scale}$,
			upper-right corner in figures \ref{fig:2}{\bf e-g}) easily become chaotic. These have the desired large
			separability (large $r^S$, figure \ref{fig:2}{\bf e}) but very low generalization capabilities (large $r^G$,
			figure \ref{fig:2}{\bf f}). This is so because of the high sensibility of chaotic systems to initial conditions.
			Performance in arbitrary tasks is best when $r^S$ and $r^G$ are relatively balanced (figure \ref{fig:2}{\bf g}).
			Reservoirs in that region of the $\lambda - \omega_{scale}$ morphospace are capable of separating relevant
			behaviors while recognizing redundancy and noise in the input data.

			The quantities $r^S$ and $r^G$ capture desirable properties of reservoir dynamics. They are also easy to measure
			empirically (see section \ref{sec:3.1}) to determine if they are any good for RC. These convenient reservoir
			properties (separability and generalization) are conflicting traits -- changing a circuit to improve one often
			degrades the performance in the other. The authors in \cite{MaassBertschinger2005, LegensteinMaass2007b,
			LegensteinMaass2007a} acknowledge that they do not have a principled way to compare $r^S$ versus $r^G$ -- figure
			\ref{fig:2}{\bf e} just shows the difference between them. We propose that Pareto optimality \cite{Coello2006,
			Schuster2012, Seoane2016} might be a well grounded framework for this problem. Pareto, or multiobjective,
			optimization is the most parsimonious way to bring together quantifiable traits that cannot be directly compared
			(e.g. because they have disparate units and dimensions), and to do so without introducing undesired biases. For
			the current problem, Pareto optimality would give us the best $r^S-r^G$ tradeoff as embodied by the subset of
			neural circuits that cannot be changed to improve both quantities simultaneously. This offers a guideline to
			select systems that perform somehow optimally towards both targets. This approach has helped identify salient
			designs in biological systems evolving under conflicting forces \cite{ShovalAlon2012, HartAlon2015,
			SzekelyAlon2015, TendlerAlon2015}. It can also link the computation capabilities of reservoirs to phase
			transitions \cite{Seoane2016, SeoaneSole2013, SeoaneSole2015a, SeoaneSole2016} and, more relevant for us,
			criticality \cite{SeoaneSole2015b}. \\

			Indeed, criticality has long been a good candidate as a governing principle of brain connectivity and dynamics
			\cite{Bak1996, BeggPlenz2003, LegensteinMaass2007a, Chialvo2010, MoraBialek2011, TagliazucchiChialvo2012,
			MorettiMunoz2013, Munoz2018}. In statistical mechanics, critical systems are rare configurations of matter poised
			between order and disorder. Such states present long-range correlations between parts of the system in space and
			time, with arguably optimal sensitivity to external perturbations. A similar phenomenon was noted in computer
			science studying cellular automata and random Boolean networks \cite{Kauffman1996, Wolfram1984, Langton1990,
			MitchellCrutchfield1993, BertschingerNatschlager2004, LegensteinMaass2007a}. Such systems usually present ordered
			and disordered phases. In the former (analogous, e.g., to solid matter in thermodynamics), activity fades away
			quickly into a featureless attractor. No memory is preserved about the initial state (which acts as computational
			input to the system) or its internal structure. The later, disordered phase presents chaotic dynamics with large
			sensitivity to the input and its inner vagaries. Slightly similar initial conditions differ quickly, thus erasing
			any correlation between potentially related inputs and resulting in trajectories without computational
			significance. Separating both behaviors lies the {\em edge of chaos}, which balances the stability of the ordered
			phase (thus building relatively lasting steady states) and the versatile dynamics of the disordered phase (which
			enables the mixing of relevant input features). These properties allow systems at the edge of chaos to optimally
			combine input parts and compute.

			This depiction of critical systems reminds us of the desirable design captured by $r^S$ and $r^G$. Several authors
			have used hallmark indicators of criticality to contrive optimal reservoirs with enhanced performance. The
			measures employed include Lyapunov exponents \cite{MaassBertschinger2005, BertschingerNatschlager2004,
			LegensteinMaass2007b, SchrauwenLegenstein2009, ToyozumiAbbott2011, BoedeckerAsada2012, BianchiAlippi2018a} and
			Fisher information \cite{BianchiAlippi2018b, LiviAlippi2018} of reservoir dynamics. The former estimates how
			rapidly slight perturbations get amplified (thus diverge) as the reservoir dynamics unfold. This divergence never
			happens if the dynamics are too ordered (perturbations get dumped, Lyapunov exponents are negative), and happens
			too quickly in chaotic regimes (positive exponents). Only at the edge of chaos (Lyapunov exponents tend to zero) a
			small perturbation can make a lasting yet meaningful difference that does not fade away or explode over time.
			Similar principles lead to diverging Fisher information as criticality is approached \cite{ProkopenkoWang2011}.

			All these works report notable correlations between such traces of criticality and enhanced reservoir performance
			-- in line with the evidence that a balanced $r^S - r^G$ is an indicator of good reservoir design
			\cite{MaassBertschinger2005, LegensteinMaass2007b, LegensteinMaass2007a}. Figure \ref{fig:2}{\bf g} also shows how
			this $r^S - r^G$ compromise degrades faster in the chaotic than in the more ordered region. Meanwhile, Toyozumi
			and Abbott \cite{ToyozumiAbbott2011} use Lyapunov exponents to suggest that reservoir performance should degrade
			faster in the ordered phase than in the disordered one. In the emerging picture, the advantages of criticality are
			clear and suggest powerful evolutionary constraints to bring naturally occurring RC-based systems towards the edge
			of chaos. But a critical state is often difficult to reach and sustain, so often we settle for getting as close as
			possible. Both \cite{MaassBertschinger2005, LegensteinMaass2007b, LegensteinMaass2007a} and
			\cite{ToyozumiAbbott2011} predict that it makes a difference from which side we approach the edge of chaos. They
			disagree on what side is computationally preferred. Empirical studies (see \cite{Munoz2018} for an up-to-date
			review) remain inconclusive too. Further research is needed.

	\section{Inspiring examples in biology and engineering} 
		\label{sec:3}

		\subsection{Reservoirs in the brain} 
			\label{sec:3.1}

			\begin{figure*}
				\begin{center}
					\includegraphics[width = \textwidth]{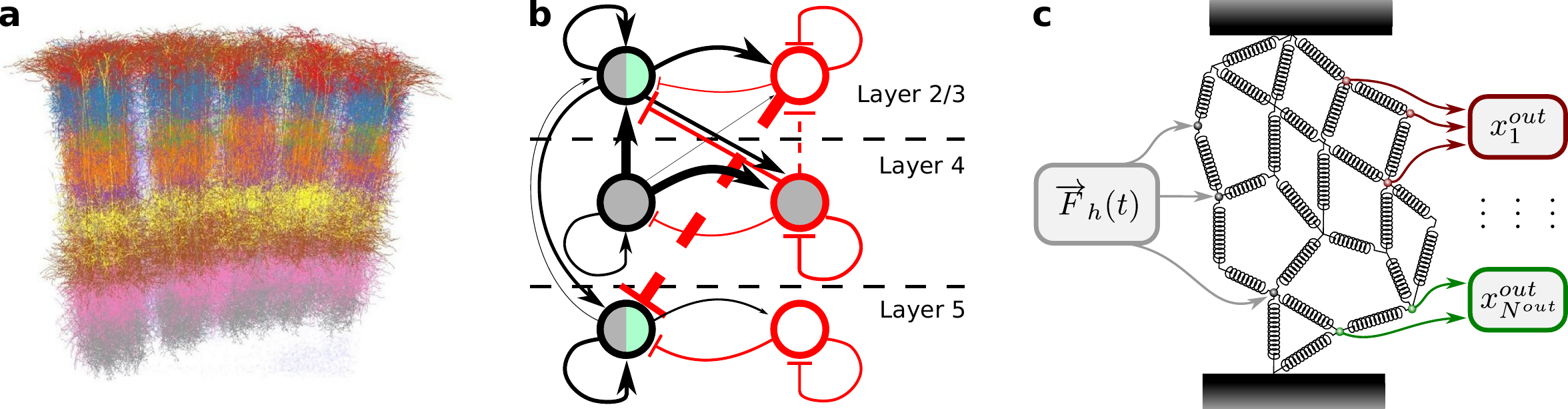}

        \caption{ {\bf Reservoirs in the brain and the body. } {\bf a} Cortical microcolumns have been proposed as the
basic operating unit of the neocortex \cite{HawkinsBlakeslee2007}. Liquid State Networks \cite{MaassMarkram2002} where
largely introduced in an attempt to clarify the computational basis of these circuits. Figure credit to Oberlaender et
al., reconstruction methods in \cite{OberlaenderSakmann2011}. {\bf b} Schematic representation of connections between
layers within a single cortical microcolumn. Figure reconstructed from \cite{HaeuslerMaass2006, HabenschussMaass2013,
Maass2016}. Black node contour and black arrows represent respectively excitatory neurons and connections, red indicates
inhibitory ones. Arrow width is proportional to connection strength as estimated from empirical data. Dashed arrows
indicate data is less significative. Gray-filled nodes indicate they receive input from other areas -- mostly cortex and
thalamus. Green-filled nodes indicate they act as output -- mostly to other cortical areas (both output nodes) and
thalamus (layer $5$ output node). {\bf c} Networks of springs can work very efficiently as reservoirs, as long as they
display heterogeneous dynamics as a response to inputs. This opens a huge potential in robotics and elsewhere, since
multiple mechanoelastical systems can be modeled as networked spring-mass structures. }

					\label{fig:3}
				\end{center}
			\end{figure*}

			When conceiving Liquid State Machines, Maass et al. drew inspiration from the cortical microcolumn (figure
			\ref{fig:3}{\bf a}). Through RC, they proposed a plausible computing strategy that could be the basic operating
			principle of these circuits \cite{MaassMarkram2002, MaassMarkram2004a, MaassMarkram2004b, MaassMarkram2006}. These
			neural motifs are roughly cylindrical structures that convey information mostly inwards from and perpendicularly
			to the neocortex surface. Contiguous columns are saliently distinguished from each other, but sparse lateral
			connections do exist. An important part of the internal structure, as well as external connections to and from
			other columns and other parts of the brain (notably the thalamus), appears to be stereotypical
			\cite{ThomsonBannister2002}. Specific connections and circuit motifs correlate with the different cortical layers
			within each column (figure \ref{fig:3}{\bf b}). Maass et al. built upon this known average structure when
			designing realistic reservoirs. Still, columns vary morphologically and functionally throughout the cortex. Most
			of them can hardly be associated to exclusive functionality, while others can be linked down to specific
			computations. For example, the receptive field of columns in V1 in cats was early identified by Hubel and Wiesel
			as responding to visual gratings with specific inclination \cite{HubelWiesel1962}; the barrel cortex in mice
			consists of cortical columns that have grown in size through evolution and have specialized in processing the
			sensory stimuli from individual whiskers \cite{DiamondAhissar2008}. It has been proposed that the cortical
			microcolumn constitutes the operational unit of the neocortex, and that the advanced cognitive success of mammal
			brains is a consequence of the exhaustive use of this versatile circuit \cite{HawkinsBlakeslee2007}. Evidence
			about this remains inconclusive. In any case, discerning the computational basis of cortical circuits is a
			relevant, open question in neuroscience to which RC can contribute greatly.

			In \cite{Maass2016}, Maass notes the heterogeneity of neuron types in the brain (also within cortical columns).
			They display great morphological diversity, varying number of presynaptic neurons, and different physiological
			constituency that results in diversified time and spacial integration scales. The recursive nature of most neural
			circuits is also highlighted. This diversity and recursivity poses a challenge to the mostly feed-forward
			computing paradigms engineered into our computers, which also rely on a relative homogeneity of its components to
			make hardware modular and reprogramable. What could then be the computational basis of neural circuits? RC comes
			in to take advantage of these features, and steps forward as a likely computational foundation of the brain.
			Looking for empirical evidence, three hallmarks have been sought in cortical circuits that would be indicative of
			RC: i) Records of this morphological heterogeneity, which in turn results in the dynamical diversity proper of a
			reservoir. ii) Existence of parallel neurons that gather information from a same processing center (akin to a
			repository) and project their output into distinct areas that solve different tasks; just as RC readouts use a
			same reservoir without interference. iii) The ability to retrieve relevant (including highly non-linear) input
			features by training simple linear classifiers on neural recordings. Note that all these would constitute rather
			circumstantial evidence for RC since each of these features could be exploited by other computing paradigms too.
			As of today, this is the best that we can do as far as signs of RC in the brain goes. We review a series of
			studies showing such indirect evidence in the next paragraphs.
			
			Neural heterogeneity resulting in desirable, reservoir-like properties has been reported in cortical neurons, the
			retina and the primary visual cortex, and in networks grown in-vitro from dissociated neurons
			\cite{HaeuslerMaass2006, BernacchiaWang2011, NikolicMaass2009, MarreBerry2015, DraniasVanDongen2013,
			JuVanDongen2015}. The existing diversity of neural components results in very heterogeneous dynamics
			\cite{Singer2013} -- which can be exploited by RC-like computation. The emergence of such dynamical heterogeneity
			in dissociated neurons reveals that this property follows spontaneously from the wiring mechanisms. Indeed,
			computational studies show how input-driven Spike-Timing Dependent Plasticity can generate heterogeneous circuitry
			that then displays good reservoir behavior \cite{KlampflMaass2013}.

			As argued above, by restricting training to the output units, RC facilitates the use of a same reservoir to solve
			different tasks by just plugging parallel readouts that do not interfere with each other. This principle seems to
			be exploited by the brain. In \cite{ChenHelmchen2013}, neurons are recorded which project from the mouse barrel
			cortex to other sensory and motor centers. Both kinds of neurons retrieve the same information, but they respond
			differently, with task specificity depending on whether further sensory processing is required or whether the
			motor system needs to be involved. From the RC framework, the barrel cortex would be working as a reservoir whose
			activity is not enough to determine which neurons will respond and how, suggesting independent wiring for each
			individual task based on a same, shared input.

			All these works casually suggest that advanced neural structures are indeed using some of the RC principles. More
			compelling evidence comes from the computational analysis of neural populations as they relate to input signals.
			More precisely, from the ability to retrieve non-linear input features by using just linear combinations of
			sparse, recorded neural activity \cite{KlampflMaass2012, RigottiFusi2013, FusiRigotti2016}. In
			\cite{KlampflMaass2012} it is shown how sparse recordings from the primary auditory cortex of ferrets (involving
			just between $4$ and $10$ neurons) are enough to retrieve non-linear features of auditory stimulus in a task that
			involves tones that increase or decrease randomly by octaves. This information can be extracted by simple linear
			classifiers trained upon the recorded neural activity. This is possible because the recorded neurons (which act as
			a reservoir) already implement a sufficient, non-linear transformation of the input. More complicated methods
			(e.g. Support Vector Machines) do not show a relevant performance increase. We can parsimoniously assume that
			evolution would settle for simpler solutions (i.e. linear readouts) if they suffice -- unless unknown selective
			pressures existed.

			The term {\em mixed selectivity} is used in \cite{RigottiFusi2013, FusiRigotti2016} to underscore that neurons
			from the reservoir in these experiments do not respond to simple (i.e., somehow linear) features of the stimulus.
			Measures akin to the $r^S$ and $r^G$ introduced above are computed in \cite{RigottiFusi2013, FusiRigotti2016} for
			a set of neural recordings while monkeys perform a series of tasks. It is shown how the classification accuracy
			grows with the dimensionality of the space (which corresponds to larger $r^S$) into which the neural recordings
			project the input stimuli. The underlying reason is, again, that this larger $r^S$ allows more different binary
			classifiers to be allocated among the data.

			The baseline story is that biological neural systems in the neocortex (and potentially other parts of the brain)
			exhibit all the ingredients needed to implement RC. Most importantly, a lot of meaningful information can be
			retrieved from real neural activity using simple linear readouts. From an evolutionary point of view, it would
			feel suboptimal to perform further complicated operations (specially provided that they do not improve performance
			\cite{KlampflMaass2012, RigottiFusi2013}).

		\subsection{The body as a reservoir}
			\label{sec:3.2}

			In \cite{HauserMaass2011}, Hauser et al. implement a reservoir using two-dimensional networks of springs (figure
			\ref{fig:3}{\bf c}). Inputs are provided as horizontal forces that displace some springs from their resting
			states. Such perturbations propagate through the network similarly to activity in other reservoirs. Simple linear
			readouts can be trained to pick up both vertical and horizontal elongations (these would constitute the internal
			state $\bar{x}(t)$ of the system), and thus perform all kind of computations upon the input signals. It seems
			trivial that such a reservoir will work as long as the springs present a variety of elastic constants (hence
			providing the richness of dynamics that RC demands). But a more important conceptual point is made in
			\cite{HauserMaass2011}: the possibility that bodies can function as reservoirs, with springs modeling muscle
			fibers and other sources of mechanical tension.

			A more explicit implementation is explored in \cite{NakajimaPfeifer2013a, NakajimaPfeifer2013b} where the muscles
			of an octopus arm are simulated and used as a reservoir. Torques at the base of the arm serve as inputs. These
			forces propagate along the arm, perturbing modules of coupled springs (figure \ref{fig:3}{\bf d}). A simple linear
			classifier reads the elongation of the various springs. The readouts are trained using standard RC methods until
			they reproduce a desired function of the output. Alternatively, readout activity is fed back to the arm and
			trained so that it displays a target motion. Octopuses have a central brain with $\sim50$ million neurons versus a
			distributed nervous system with $\sim300$ million neurons \cite{NakajimaPfeifer2013a}. The computational power of
			nerve cells along the arms is beyond doubt \cite{GodfreySmith2016, Sacks2017}. But this approach is telling us
			something much more important: a lot of the non-linear calculations needed to process and control an arm's motion
			could be provided for free by spurious mechanical forces picked up by simple linear classifiers.

			The non-linearities and unpredictable behavior of soft tissue could have been a nuisance in robotics. They could
			have been perceived as untameable systems, very costly to simulate, that a central controller would need to
			oversee in real time. But a recent trend termed {\em morphological computation} \cite{PfeiferBongard2006,
			PfeiferIida2007, MullerHoffman2017} exploits these  non-linearities, self-organization, and in general the ability
			that soft tissues and compliant elements have shown to carry out complex computations. This framework includes
			simple behaviors such as passive walkers \cite{McGeer1990, WisseVanFrankenhuyzen2006}, materials optimized to
			provide sensory feedback \cite{FendPfeifer2006}, or collective self-organization of smaller robots
			\cite{MurataKurokawa2007}. The approach of the body as a reservoir (demonstrated by the networks of springs just
			described \cite{HauserMaass2011, NakajimaPfeifer2013a, NakajimaPfeifer2013b, SumiokaPfeifer2011, HauserMaass2012,
			NakajimaPfeifer2013c} or by tensegrity structures that can crawl controlled by RC-based feedback
			\cite{CaluwaertsSchrauwen2011, CaluwaertsSchrauwen2013}) offers a principled way to develop a sound theory of
			morphological computation \cite{HauserMaass2011, HauserMaass2012, MullerHoffman2017}.

			Other bodily elements besides physical tensions can work as a reservoir. Recently, Gabalda-Sagarra et al.
			\cite{GabaldaSagarraGarciaOjalvo2018} have shown how the Gene Regulatory Network (GRN) in a range of cells (from
			bacteria to humans) present a structure quite suited for RC. Empirically-inspired GRNs are simulated and used as
			reservoirs to solve benchmark problems as good as known optimal topologies. They also show how an evolutionary
			process could successfully train output readouts stacked on top of those GRNs. These examples with physical bodies
			and gene cross-regulation highlight a potential abundance of repertoires in Nature.

			They also highlight the importance of {\em embodied computation} \cite{PfeiferIida2007, Clark1998} -- the fact
			that living systems develop their behavior within a physical reality whose elements (including bodies) can
			participate in the needed calculations, become passive processors, expand an agent's memory, etc. In robotics,
			this opens up huge possibilities \cite{NakajimaPfeifer2014, NakajimaPfeifer2015} -- e.g., to outsource much of the
			virtual operations needed to simulate robot bodies. From an evolutionary perspective, the powerful and affordable
			computations that RC offers through compliant bodies raises a series of questions. For example, with animal motor
			control in mind: Since RC is a valid approach to the problem, and it seems to provide so much computational power
			for free, why is it not more broadly used? Why would, instead, a centralized model and simulation of our body (as
			the one harbored by the sensory-motor areas of the cortex) become so prominent instead? What were the evolutionary
			forces shaping this process, which somehow displaced computation from its embodiment to favor a more {\em virtual}
			approach? It is still possible that, unknown to us, RC actually takes place with the body (or parts of the body)
			as a reservoir -- after all, the paradigm has only been introduced recently (and already see
			\cite{ShimHusbands2007, ValeroCuevasLipson2007, MullerHoffman2017} for examples falling close enough). However,
			the most salient features of advanced motor control (e.g., sensory-motor cortices, the central pattern generator
			that regulates gait, some peripheral circuits implementing reflexes) do not resemble RC much. So the above
			questions can still teach us something about how RC endures different selective pressures. A possibility that we
			explore further in the next section is that RC shall be an unstable evolutionary solution.

	\section{Discussion -- evolutionary paths to Reservoir Computing}
		\label{sec:4}

		RC is a very cheap and versatile paradigm. By exploiting a reservoir capable of extracting spatiotemporal, 
		non-linear features from arbitrary input signals, simple linear classifiers suffice to solve a large collection of
		tasks including classification, motor control, time-series forecasting, etc \cite{VerstraetenStroobandt2006,
		JaegerSiewert2007, SoriaRuffini2018, JoshiMaass2004, SalmenPloger2005, Burgsteiner2005, JaegerHaas2004,
		IbanezSoriaRuffini2018, VerstraetenVanCampenhout2005, TongCottrell2007, TriefenbachMartens2010, HinautDominey2013}.
		This approach simplifies astonishingly the problem of training Recurrent Neural Networks, a job plagued with hard
		numerical and analytic difficulties \cite{BengioFrasconi1994, PascanuBengio2018}. Furthermore, as we have seen,
		reservoir-like systems abound in Nature: from non-linearities in liquids and GRNs \cite{FernandoSojakka2003,
		GabaldaSagarraGarciaOjalvo2018}, through mechanoelastic forces in muscles \cite{HauserMaass2011,
		NakajimaPfeifer2013a, NakajimaPfeifer2013b}, to the electric dynamics across neural networks \cite{MaassMarkram2002,
		MaassBertschinger2005, LegensteinMaass2007b}; a plethora of systems can be exploited as reservoirs. Reading off
		relevant, highly non-linear information from an environment becomes as simple as plugging linear perceptrons into
		such structures. Adopting the RC viewpoint, it appears that Nature presents a trove of meaningful information ready
		to be exploited and coopted by Darwinian evolution or engineers so that more complex shapes can be built and ever
		more intricate computations can be solved.

		When looking at RC from an evolutionary perspective these advantages pose a series of questions. Where and how is RC
		actually employed? Why is this paradigm not as prominent as its power and simplicity would suggest? In biology, why
		is RC not exploited more often by living organisms (or is it?); in engineering, why is RC only so recently making a
		show. This section is a speculative exercise around these points. We will suggest a series of factors which, we
		think, are indispensable for RC to emerge and, more importantly, to persist over evolutionary time. Based on these
		factors we propose a key hypothesis: while RC shall emerge easily and reservoirs abound around us, these are not
		evolutionarily stable designs as systems specialize or scale up. If reservoirs evolve such that signals need to
		travel longer distances (e.g. over bigger bodies), integrate information from senses with wildly varying time
		scales, or carry out very specific functions (such that the generalizing properties of the reservoir are not needed
		anymore), then the original RC paradigm might be abandoned in favor of better options. Then, fine-tuned, dedicated
		circuits might evolve from the raw material that reservoirs offer. A main goal of this speculative section is to
		provide testable hypotheses that can be tackled computationally through simulations, thus suggesting open research
		questions at the interface between computation and evolution.

		First of all, we should not dismiss the possibility that RC has been overlooked around us -- it might actually be a
		frequent computing paradigm in living systems. It has only recently been introduced, which hints us that it is not
		as salient or intuitive as other computing approaches. There was a lot of mutual inspiration between biology and
		computer science as perceptrons \cite{Minsky2017}, attractor networks \cite{Hopfield1982}, or self-organized maps
		\cite{Kohonen1982} were introduced. Prominent systems in our brain clearly seem to use these and other known
		paradigms \cite{Hebb1963, FukushimaMiyake1982, YaminsDiCarlo2014, KhalighRazaviKriegeskorte2014, Ritter1990,
		SpitzerKischka1995}. We expect that RC is used as well. We have reviewed some evidence suggesting that it is
		exploited by several neural circuits \cite{HaeuslerMaass2006, BernacchiaWang2011, NikolicMaass2009, MarreBerry2015,
		DraniasVanDongen2013, JuVanDongen2015, ChenHelmchen2013, KlampflMaass2012, RigottiFusi2013, FusiRigotti2016}, or by
		body parts using the morphological computation approach \cite{ValeroCuevasLipson2007, ShimHusbands2007}. All this
		evidence, while entailing, is far from, e.g., the visually appealing similarity between the structure of the visual
		cortices and modern, deep convolutional neural networks for computer vision \cite{KrizhevskyHinton2012,
		YaminsDiCarlo2014, KhalighRazaviKriegeskorte2014} (figure \ref{fig:1}{\bf e}). Altogether, it seems fair to say that
		RC in biology is either scarce or elusive, even if only recently we are looking at biological systems through this
		optic.

		\begin{figure*}
			\begin{center}
				\includegraphics[width = \textwidth]{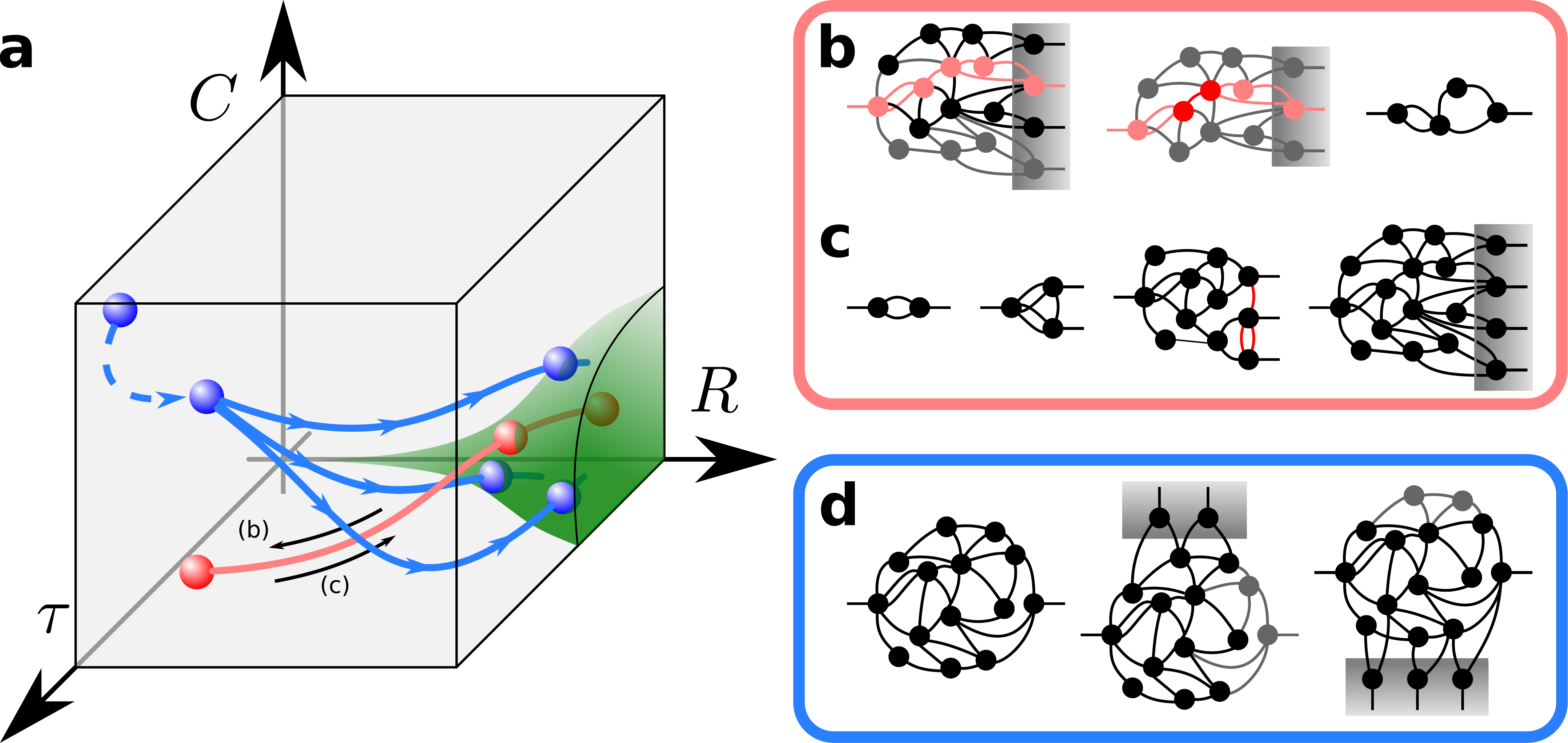}

        \caption{ {\bf Evolutionary paths to and from RC. } {\bf a} We propose a morphospace to locate RC-based circuits
(this could be extended to other computing paradigms). The axes are determined by the circuit cost ($C$), and two
aspects of an underlying fitness landscape that drives circuit selection: its {\em ruggedness} ($R$) and the average
lifetime ($\tau$) that computational tasks contribute to fitness. RC should ensue when both $C$ and $\tau$ are low and
$R$ is high (green volume). Low $\tau$ and high $R$ demand, respectively, cheap learning and multitasking. Possible
evolutionary paths of non-RC circuits into the RC area (and vice-versa) are depicted. {\bf b} (Red curve in panel {\bf
a}.) Readout units implement simple linear classifiers (indicated by the gray rectangle). This RC-based circuit has
discovered a very fit computation implemented by the red units. The other readouts and their tasks (as well as some
components of the reservoir) gradually become spurious and evolutionary dynamics get rid of them. Eliminating some nodes
can prompt a reorganization of the original units implementing the valuable computation (solid red units). Such a
circuit migrates through the morphospace from the RC region into a position with little ruggedness and stable peaks of
the landscape. The cost, too, should have been decreased. {\bf c} (Also red curve in panel {\bf a}, traversed in the
opposite direction.) A simple circuit gradually becomes more complex. New tasks are implemented as components are
appended. If true RC is reached, feedback between the outputs (marked with red links) should be lost and readout units
should implement linear classifiers (again, gray rectangle). At some point, this should be noted as a certain symmetry
breaking between reservoir and readout units. This might leave an empirical trace in biological systems. {\bf d} (Blue
trajectories in panel {\bf a}.) A genetic mutation produces two copies of a very specialized circuit. One of the copies
keeps implementing its very specific task (blue ball in the upper left corner of the morphospace, panel {\bf a}). This
represents a lasting, singled out peak in fitness landscape and the implementation is costly because mistakes would
cause a fatal failure. The circuit copy is released from the cost of failure (jump represented by the blue arrow in
panel {\bf a}). It can then wander the morphospace freely (different blue trajectories), probably becoming cheaper as
some further functionality is lost. Its very rich dynamics make it a perfect candidate to be hijacked by simple, linear
readouts that start using the circuit as a reservoir. }

				\label{fig:4}
			\end{center}
		\end{figure*}
		
		The two main advantages brought about by RC are: i) very cheap learning and ii) a startling capability for parallel
		processing. Its main drawback compared to other paradigms is the amount of extra activity needed to capture
		incidental input features that might never be actually used. We can view these aspects of RC as evolutionary
		pressures defining the axes of a morphospace. Morphospaces are an insightful picture that has been used to relate
		instances of natural \cite{Raup1966, Niklas1997, McGhee1999, Niklas2004} and synthetic
		\cite{CorominasMurtraRodriguezCaso2013, AvenaKoenigsbergerSporns2015, SeoaneSole2018b} complex systems to each other
		guided by metrics (sometimes rigorous, other times qualitative) that emerge from mathematical models or empirical
		data. Here we lean towards the qualitative side, but it should also be possible to qualitatively locate RC and other
		computational paradigms in the morphospace that follows. That would allow us to compare these different paradigms,
		or different circuit topologies within each paradigm, against each other under evolutionary pressures.

		A first axis is straightforwardly the {\em dynamical cost} ($C$, figure \ref{fig:4}{\bf a}) of the reservoir, since
		RC demands so much more activity than it eventually uses. This would prevent RC at large organism scales with costly
		metabolism, but still allows myriad smaller physical systems (such as muscles or tiny bodies) to behave as {\em
		free-floating} reservoirs ready to be exapted.

		Which brings in the next question: given these freely available reservoirs (specifically supported by the spring and
		gene regulatory network examples), when will they be exploited? To answer this, let us focus on the two main
		advantages of RC mentioned above, starting with the cheap learning. Let us also assume that there exists an
		underlying fitness landscape that tells us whether a feature (e.g. solving a specific computation) contributes to
		the success of a living organism. We conceive learning as a process much faster than evolutionary dynamics. Looking
		at the fitness landscape, we would expect features that offer fitness over long evolutionary periods to be 
		hard-wired, not learned. We would not need a reservoir to capture these; but rather a robust, efficient, and
		dedicated structure fixed by evolution. Features which are rather learned, on the other hand, shall offer fitness on
		a time-scale briefer than a lifetime. We are talking about short-lived peaks of the landscape, so voluble or
		unpredictable that it becomes preferable to keep a learning engine rather than hardwiring a fixed design. Some
		notion of an {\em average lifetime} ($\tau$, figure \ref{fig:4}{\bf a}), during which a feature contributes to
		fitness, defines a second axis of our morphospace.

		Similarly, to exploit the parallel processing abilities of RC, the underlying fitness landscape should be peaked
		with multiple optima that represent different useful computations. Thus {\em ruggedness} ($R$, figure
		\ref{fig:4}{\bf a}) defines the last axis of our morphospace. If one or few tasks contribute much more fitness than
		others, parts of the reservoir dedicated to them would be reinforced over evolutionary times and unimportant
		components would fade away, eventually thinning down the reservoir and dismissing the less fit computations (figure
		\ref{fig:4}{\bf b}). Over very short time-scales, the ruggedness and peak lifetime axes shall become
		indistinguishable: a quickly shifting single peak reminds a rugged landscape when looked at from afar.

		These wildly speculative hypotheses suggest a research program amid evolution and computation. We can craft
		artificial fitness landscapes with computing tasks at their peaks. We could then mutate and select reservoirs with
		costs associated to their dynamics, and with a reward collected as they solve tasks around the shifting landscape.
		We expect RC to fade away if some of the conditions are removed. For example, if a reservoir grows in physical size
		so that communicating dynamical states over long distances becomes metabolically costly. Might this have happened in
		motor control for larger bodies? How would be the interplay between the advantages afforded by RC and a growing
		morphology? What would be the evolutionary fate of the components of a reservoir? Under what conditions do
		reservoirs retain their full original architecture with redundant dynamics? When do they thin down to a subset of
		dedicated parts that capture specific signals ignoring the rest (figure \ref{fig:4}{\bf b})? Could some conditions
		prompt the development of more complex reservoirs? Can we observe a reservoir complexity ratchet -- a threshold
		beyond which RC becomes evolutionary robust? These are all issues that can be tackled through simulations in a
		systematic and easy manner. Also, these questions extend beyond RC. We used its properties as guidelines to design
		morphospace axes that, we think, could optimally tell apart this computing strategies from others. But both the
		lifetime and ruggedness of tasks in our landscape are independent of the strategy used to tackle each problem. A
		dynamical cost can also be calculated in other computing devices. Hence, we expect that the morphospace will be
		useful in locating RC among a larger family of RNN and other paradigms. The eventual picture could contain hybrids
		as well, thus potentially revealing a continuous of computing options. 

		Back to RC, up to this point we have assumed that a full-fledged reservoir exists which, depending on external
		constraints, might shift to simpler computing paradigms and lose part of its structure (figure \ref{fig:4}{\bf b}).
		This shall be relevant for the kind of reservoirs provided for free by Nature, as discussed above. But there are
		alternative questions at the other end of the spectrum: What would be plausible evolutionary paths for non-RC
		circuits to progress towards RC? If cortical microcolumns actually implement RC in our brains, we could then wonder
		how they got there by building upon non-RC elements. Some non-exhaustive possibilities include:
			\begin{itemize}

        \item Gradually, over evolutionary time, a small, specialized circuit acquires more tasks at the same time that
it becomes more complex -- e.g. by incorporating more parts and a more convoluted topology (figure \ref{fig:4}{\bf c}).
For this to result in RC, partial computations needed for the acquired tasks must be distributed over (somehow
involving) several circuit components in a way that makes them difficult to disentangle. Otherwise, the system would
more likely evolve into smaller, separate, and task- specific structures. Importantly, at some point, a kind of
symmetry-breaking between the reservoir and the readouts should take place. All the costly, non-linear calculations
should be relegated to the parts that will constitute the reservoir. Readouts or circuit effectors, on the other hand,
can become simpler -- with linear classifiers eventually sufficing to do the job. Given the way how learning works in
RC, which only affects the readouts, this symmetry breaking between parts should also be reflected by a specialization
of feedbacks controlling synaptic plasticity -- a point that we will discuss again below.

        \item An already complex, yet specialized circuit is freed of its main task, thus becoming raw material that can
be coopted to solve other computations (figure \ref{fig:4}{\bf d}). This evolutionary path to RC could be explored,
e.g., if a complex circuit gets duplicated and one of the copies keeps implementing the original task -- so that
outlandish explorations are not penalized. This is a mechanism exploited elsewhere in biology, e.g., by duplicated genes
with one of the copies exploring the phenotypic neighborhood of an existing peak in fitness landscape. (Indeed, we could
look at such pools of duplicated genes as a reservoir of sorts; and we could wonder whether the structure, variability,
and frequency of such {\em gene reservoirs} could help us quantify aspects of our RC morphospace.) The similarities
between Central Pattern Generators in the brain stem and columns in the neocortex have been noted at several levels
including dynamical, histological, biomolecular, patological, and structural \cite{YusteLansner2005}. A key difference
is the versatility and plasticity of microcolumns when compared with the sterotypical behavior of CPGs (a versatility
that would be very costly for CPGs, since they would fail to implement their main task). RC is explicitly suggested in
\cite{YusteLansner2005} as a paradigm to frame the differences between CPGs and cortical columns. The hypothesis that
microcolumns shall have followed this path to RC from a shared evolutionary origin with CPGs becomes tempting.

			\end{itemize}

		These paths to RC, again, work as evolutionary-computational hypotheses that can be easily tested through
		simulations. It is perhaps also possible to derive some mean-field models of average circuit structure and their
		computational power and address some limit cases analytically. These are all engaging topics for the future, but a
		look at the foundation of our speculations already offers hints to answer some of the questions above -- notably,
		why is RC not a more prominent paradigm.

		While circuits and complex systems with the ability to work as reservoirs abound in Nature, situations that sustain
		evolutionary pressures with a rugged, sifting landscape (demanding multitasking and adaptability within times much
		shorter than the lifetime of the reservoir) might not be as common. A possibility is that a peak of the fitness
		landscape becomes more prominent for an evolving species -- e.g., by a process of niche construction. Then, a single
		task among the several ones solved by a reservoir can provide enough fitness. Redundant tasks and components can be
		consequently lost (figure \ref{fig:4}{\bf b}). As suggested in the previous section, against the convenience of
		bodies as a reservoir, we propose that this might be a reason why nerves at large evolved towards a more sequential
		and archetypal wiring. As bodies grew bigger, peaks in the motor-control landscape might have become more prominent
		(e.g., just a few coarse-grained commands seem enough for CPGs to coordinate gait behavior at large
		\cite{MarderCalabrese1996, Ijspeert2008}). Even though bodies as reservoirs still serve the information needed to
		solve motor control, the archetypal circuitry is more stable in the long run. The need to integrate visual cues
		(which can hardly be incorporated into a mechanoelastic reservoir) and long-term movement planning hint at yet other
		evolutionary pressures against the RC solution. This suggests that smaller or more primitive organisms, if any,
		shall exploit RC more clearly. Just as RC appears as a relevant tool to develop a principled theory of morphological
		computation \cite{HauserMaass2011, HauserMaass2012, MullerHoffman2017}, it also seems a great addition to think
		about Liquid and Solid Brains along the lines of this volume, especially as embodied and non-standard computations
		with small living organisms are explored \cite{BaluskaLevin2016}.

		Cortical microcolumns, on the other hand, are likely to operate based on RC or to incorporate most RC principles.
		These are the circuits on top of the more abstract and complex tasks, such as language or conscious processing.
		These tasks appear indeed shifting in nature, and often presenting a wide variety of solutions (e.g. as indicated by
		the different syntax and grammars that implement language equally well). These features loosely correspond to rugged
		landscape whose peaks either shift in time or cannot be anticipated from the long evolutionary perspective -- thus
		fitting nicely in our speculation. Efforts to clarify whether RC is actually exploited in the neocortex and other
		neural circuits is underway, as reviewed earlier \cite{HaeuslerMaass2006, BernacchiaWang2011, NikolicMaass2009,
		MarreBerry2015, DraniasVanDongen2013, JuVanDongen2015, ChenHelmchen2013, KlampflMaass2012, RigottiFusi2013,
		FusiRigotti2016}. These works focus mostly on the richness of the dynamics and on the ability of simple linear
		classifiers to pick up non-linear input features based on recorded neural activity alone.

		We would like to add an alternative empirical approach: As mentioned above, RC implies that training focuses on the
		linear readouts. This must be reflected at several levels. As an instance, the target behavior must be made
		available to the readouts (e.g., to implement back propagation or Hebbian learning), but is not needed elsewhere. On
		the other hand, we have explored a series of computation-theoretical features that reservoirs should preferably
		display. These include a tendency to criticality and a simultaneous maximization of the separability and
		generalization properties (as measured by $r^S$ and $r^G$). All these pose diverging evolutionary targets for
		reservoir and readout plasticity. This should result in differences in the mechanisms guiding neural wiring --
		perhaps even at the molecular level. Trying to spot such differences empirically should be within the reach of
		current technology. \\

		We close with a reflection to link this paper with the general issue of this volume. In the introduction we
		anticipated that most authors would be exploring liquid or solid brains as referred to the thermodynamic state of
		the physical substrate in which computation happens -- i.e. whether computing units are motile (ants, T-cells, ...)
		or fixed (neurons, semiconductors, ...). Instead, this paper rather tackled aspects of the software. RC largely
		resembles a liquid in the signals involved (sometimes literally so \cite{FernandoSojakka2003}), specially through
		the abundance of spurious dynamics elicited. Evolutionary costs could limit these generous dynamics so that RC is
		lost (e.g., figure \ref{fig:4}{\bf b}). This could somehow {\em crystallize} the available {\em liquid} signals to a
		handful of stereotypical patterns. This could, in turn, either lower the demands on the hardware (as it requires
		less active dynamics), and otherwise free resources that could be invested, e.g., in hardware motility. We expect
		non-trivial interplays between afforded {\em liquidity} at the software and hardware levels. It then becomes
		relevant to determine whether and how the evolutionary pressures discussed above can constrain the {\em liquid} or
		{\em solid} nature of brain substrates, not only of its signal repertoire.

	\section*{Acknowledgments}
		
		We would like to thank members of the CSL for useful discussion, especially Prof. Ricard Sol\'e, Jordi Pi\~nero, and
		Blai Vidiella, as well as Dr. Amor from the Gore Lab at MIT's Physics of Living Systems. We also would like to thank
		all participants of the `Liquid Brains, Solid Brains' working group at the Santa Fe Institute for their insights
		about computation, algorithmic thinking, and biology.

	\vspace{0.2 cm}

	\section*{Competing interests}

		We declare no competing interests. 

	\vspace{0.2 cm}

	\section*{Funding} 

    This work has been supported by the Bot\'in Foundation, by Banco Santander through its Santander Universities Global
Division, a MINECO FIS2015-67616 fellowship, and the Secretaria d'Universitats i Recerca del Departament d'Economia i
Coneixement de la Generalitat de Catalunya.


\vspace{0.2 cm}

	\begin{thebibliography}{99}

		\bibitem{SzathmaryMaynardSmith1997}
			Szathm\'ary E, Maynard-Smith J. 1997 
			From replicators to reproducers: the first major transitions leading to life. 
			J. Theor. Biol. 187, 555-571.

		\bibitem{Joyce2002}
			Joyce GF. 2002 
			Molecular evolution: booting up life. 
			Nature 420(6913), 278.

		\bibitem{WalkerDavies2013}
			Walker SI, Davies PC. 2013 
			The algorithmic origins of life. 
			J. R. Soc. Interface 10(79), 20120869.

		\bibitem{Schuster1996}
			Schuster P. 1996 
			How does complexity arise in evolution: Nature's recipe for mastering scarcity, abundance, and unpredictability. 
			Complexity 2(1), 22-30.

		\bibitem{Smith2000}
			Smith JM. 2000 
			The concept of information in biology. 
			Philos. Sci. 67(2), 177-194.

		\bibitem{JablonkaLamb2006}
			Jablonka E, Lamb MJ. 2006 
			The evolution of information in the major transitions. 
			J. Theor. Biol. 239, 236-246. 

		\bibitem{Nurse2008}
			Nurse P. 2008 
			Life, logic and information. 
			Nature 454(7203), p.424.

		\bibitem{Joyce2012}
			Joyce GF. 2012 
			Bit by bit: the Darwinian basis of life. 
			PLoS Biol. 10(5), e1001323.

		\bibitem{Adami2012}
			Adami C. 2012 
			The use of information theory in evolutionary biology. 
			Ann. NY Acad. Sci. 1256(1), 49-65.

		\bibitem{HidalgoMaritan2014}
			Hidalgo J, Grilli J, Suweis S, Mu\~noz MA, Banavar JR and Maritan A. 2014 
			Information-based fitness and the emergence of criticality in living systems. 
			Proc. Nat. Acad. Sci. 111(28), 10095-10100.

		\bibitem{SmithMorowitz2016}
			Smith E, Morowitz HJ. 2016 
			The origin and nature of life on Earth: the emergence of the fourth geosphere. 
			Cambridge University Press.

		\bibitem{Hopfield1994}
			Hopfield, JJ. 1994 
			Physics, computation, and why biology looks so different. 
			J. Theor. Biol. 171(1), 53-60.

		\bibitem{Jacob1998}
			Jacob F. 1998 
			{\em Of flies, mice and man}. 
			Harvard, MA: Harvard University Press.

		\bibitem{Wagensberg2000}
			Wagensberg J. 2000 
			Complexity versus uncertainty: the question of staying alive. 
			Biol. Phil. 15, 493-508. 

		\bibitem{SeoaneSole2018a}
			Seoane LF, Sol\'e R. 2018 
			Information theory, predictability and the emergence of complex life. 
			Roy. Soc. Open Sci. 5(2), 172221.

		\bibitem{PaunSalomaa2005}
			Paun G, Rozenberg G, Salomaa A. 2005 
			DNA computing: new computing paradigms. 
			Springer Science \& Business Media.

		\bibitem{Doudna2017}
			Doudna JA, Sternberg SH. 2017 
			A crack in creation: Gene editing and the unthinkable power to control evolution. 
			Houghton Mifflin Harcourt.

		\bibitem{Thomas1973}
			Thomas R. 1973 
			Boolean formalization of genetic control circuits. 
			J. Theor. Biol. 42(3), 563-585.

		\bibitem{Kauffman1996}
			Kauffman S. 1996 
			At home in the universe: The search for the laws of self-organization and complexity. 
			Oxford university press.

		\bibitem{RodriguezCasoSole2009}
			Rodr\'iguez-Caso C, Corominas-Murtra B, Sol\'e R. 2009 
			On the basic computational structure of gene regulatory networks. 
			Mol. Biosyst. 5(12), 1617-1629.

		\bibitem{DayanAbbott2001}
			Dayan P, Abbott LF. 2001 
			Theoretical neuroscience. 
			Cambridge, MA: MIT Press.

		\bibitem{Seung2012}
			Seung S. 2012 
			Connectome: How the brain's wiring makes us who we are. 
			HMH.

		\bibitem{Levick1967}
			Levick WR. 1967 
			Receptive fields and trigger features of ganglion cells in the visual streak of the rabbit's retina. 
			J. Physiol. 188(3), 285-307.

		\bibitem{RussellWerblin2010}
			Russell TL, Werblin FS. 2010 
			Retinal synaptic pathways underlying the response of the rabbit local edge detector. 
			J. Neurophysiol. 103(5), 2757-2769.

		\bibitem{MarrHildreth1980}
			Marr D, Hildreth E. 1980 
			Theory of edge detection. 
			Proc. R. Soc. Lond. B 207(1167), 187-217.

		\bibitem{Marr1982}
			Marr D. 1982 
			Vision: A computational investigation into the human representation and processing of visual information. 
			MIT Press. Cambridge, Massachusetts.

		\bibitem{StephensBialek2013}
			Stephens GJ, Mora T, Tka\v{c}ik G and Bialek W. 2013 
			Statistical thermodynamics of natural images. 
			Phys. Rev. Let. 110(1), 018701.

		\bibitem{FukushimaMiyake1982}
			Fukushima K, Miyake S. 1982 
			Neocognitron: A self-organizing neural network model for a mechanism of visual pattern recognition. 
			In Competition and cooperation in neural nets (pp. 267-285). Springer, Berlin, Heidelberg.

		\bibitem{KrizhevskyHinton2012}
			Krizhevsky A, Sutskever I, Hinton GE. 2012 
			Imagenet classification with deep convolutional neural networks. 
			In Advances in neural information processing systems, 1097-1105.

		\bibitem{YaminsDiCarlo2014}
			Yamins DL, Hong H, Cadieu CF, Solomon EA, Seibert D, DiCarlo JJ. 2014 
			Performance-optimized hierarchical models predict neural responses in higher visual cortex. 
			Proc. Nat. Acad. Sci. 111(23), 8619-8624.

		\bibitem{KhalighRazaviKriegeskorte2014}
			Khaligh-Razavi SM, Kriegeskorte N. 2014 
			Deep supervised, but not unsupervised, models may explain IT cortical representation. 
			PLoS Comp. Biol. 10(11), e1003915.

		\bibitem{Jaeger2001}
			Jaeger H. 2001 
			The ``echo state'' approach to analysing and training recurrent neural networks-with an erratum note. 
			Bonn, Germany: German National Research Center for Information Technology GMD Technical Report, 148(34), 13.

		\bibitem{MaassMarkram2002}
			Maass W, Natschl\"ager T, Markram H. 2002 
			Real-time computing without stable states: A new framework for neural computation based on perturbations. 
			Neural Comput. 14(11), 2531-2560.

		\bibitem{JaegerPrincipe2007}
			Jaeger H, Maass W, Principe J. 2007 
			Special issue on echo state networks and liquid state machines.
			Neural Networks 20(3). 

		\bibitem{VerstraetenStroobandt2007}
			Verstraeten D, Schrauwen B, d'Haene M, Stroobandt D. 2007 
			An experimental unification of reservoir computing methods. 
			Neural networks 20(3), 391-403.

		\bibitem{LukoseviciousSchrauwen2012}
			Luko\v{s}evi\v{c}ius M, Jaeger H, Schrauwen B. 2012 
			Reservoir computing trends. 
			KI-K\"unstliche Intelligenz 26(4), 365-371.

		\bibitem{RumelhartWilliams1986}
			Rumelhart DE, Hinton GE, Williams RJ. 1986 
			Learning representations by back-propagating errors. 
			Nature 323(6088), p.533.

		\bibitem{BengioFrasconi1994}
			Bengio Y, Simard P, Frasconi P. 1994 
			Learning long-term dependencies with gradient descent is difficult. 
			IEEE T Neural Networks 5(2), 157-166.

		\bibitem{PascanuBengio2018}
			Pascanu R, Mikolov T, Bengio Y. 2013 
			On the difficulty of training recurrent neural networks. 
			In International Conference on Machine Learning, 1310-1318.

		\bibitem{VerstraetenStroobandt2006}
			Verstraeten D, Schrauwen B, Stroobandt D. 2006 
			Reservoir-based techniques for speech recognition. 
			In The 2006 IEEE International Joint Conference on Neural Network Proceedings (1050-1053). IEEE.

		\bibitem{JaegerSiewert2007}
			Jaeger H, Luko\v{s}evi\v{c}ius M, Popovici D, Siewert U. 2007 
			Optimization and applications of echo state networks with leaky-integrator neurons. 
			Neural Networks 20(3), 335-352.

		\bibitem{SoriaRuffini2018}
			Soria DI, Soria-Frisch A, Garc\'ia-Ojalvo J, Picardo J, Garc\'ia-Banda G, Servera M and Ruffini G. 2018 
			Hypoarousal non-stationary ADHD biomarker based on echo-state networks. 
			bioRxiv, p.271858.

		\bibitem{JoshiMaass2004}
			Joshi P, Maass W. 2004 
			Movement generation and control with generic neural microcircuits. 
			In: Biologically inspired approaches to advanced information technology 
			(Ijspeert A, Murata A, Wakamiya N, eds) 258-273. Berlin: Springer. 

		\bibitem{SalmenPloger2005}
			Salmen M, Ploger PG. 2005 
			Echo state networks used for motor control. 
			In Robotics and Automation, 2005. ICRA 2005. Proceedings of the 2005 IEEE International Conference on (1953-1958). IEEE.

		\bibitem{Burgsteiner2005}
			Burgsteiner H. 2005 
			Training networks of biological realistic spiking neurons for real-time robot control. 
			In Proceedings of the 9th international conference on engineering applications of neural networks, Lille, France (pp. 129-136).

		\bibitem{JaegerHaas2004}
			Jaeger H, Haas H. 2004 
			Harnessing nonlinearity: Predicting chaotic systems and saving energy in wireless communication. 
			Science 304(5667), 78-80.

		\bibitem{IbanezSoriaRuffini2018}
			Ib\'a\~nez-Soria D, Garc\'ia-Ojalvo J, Soria-Frisch A, Ruffini G. 2018 
			Detection of generalized synchronization using echo state networks. 
			Chaos 28(3), 033118.

		\bibitem{VerstraetenVanCampenhout2005}
			Verstraeten D, Schrauwen B, Stroobandt D, Van Campenhout J. 2005 
			Isolated word recognition with the liquid state machine: a case study. 
			Information Processing Letters, 95(6), 521-528.

		\bibitem{TongCottrell2007}
			Tong MH, Bickett AD, Christiansen EM, Cottrell GW. 2007 
			Learning grammatical structure with echo state networks. 
			Neural networks, 20(3), 424-432.

		\bibitem{TriefenbachMartens2010}
			Triefenbach F, Jalalvand A, Schrauwen B, Martens JP. 2010 
			Phoneme recognition with large hierarchical reservoirs. 
			In Advances in neural information processing systems (2307-2315).

		\bibitem{HinautDominey2013}
			Hinaut X, Dominey PF. 2013 
			Real-time parallel processing of grammatical structure in the fronto-striatal system: A recurrent network simulation study using reservoir computing. 
			PloS one, 8(2), p.e52946.

		\bibitem{MaciaSole2014}
			Macia J, Sole R. 2014 
			How to make a synthetic multicellular computer. 
			PLoS One 9(2), p.e81248.

		\bibitem{WyffelsStroobandt2008}
			Wyffels F, Schrauwen B, Stroobandt D. 2008 
			Stable output feedback in reservoir computing using ridge regression. 
			In International conference on artificial neural networks (pp. 808-817). 
			Springer, Berlin, Heidelberg.

		\bibitem{LukoseviciousJaeger2009}
			Luko\v{s}evi\v{c}ius M, Jaeger H. 2009 
			Reservoir computing approaches to recurrent neural network training. 
			Comput. Sci. Rev. 3(3), 127-149.

		\bibitem{Jaeger2002}
			Jaeger H. 2002 
			Tutorial on training recurrent neural networks, covering BPPT, RTRL, EKF and the ``echo state network'' approach (Vol. 5). 
			Bonn: GMD-Forschungszentrum Informationstechnik.

		\bibitem{Lukosevicious2012}
			Luko\v{s}evi\v{c}ius M. 2012 
			A practical guide to applying echo state networks. 
			In Neural networks: Tricks of the trade (pp. 659-686). 
			Springer, Berlin, Heidelberg.

		\bibitem{MaassSontag2007}
			Maass W, Joshi P, Sontag ED. 2007 
			Computational aspects of feedback in neural circuits. 
			PLoS Comput. Biol., 3(1), p.e165.

		\bibitem{SussilloAbbott2009}
			Sussillo D, Abbott LF. 2009 
			Generating coherent patterns of activity from chaotic neural networks. 
			Neuron, 63(4), pp.544-557.

		\bibitem{DaiHarley2009}
			Dai J, Venayagamoorthy GK, Harley RG. 2009 
			An introduction to the echo state network and its applications in power system. 
			In Intelligent System Applications to Power Systems, 2009. 
			ISAP'09. 15th International Conference on (pp. 1-7). IEEE.

		\bibitem{RivkindBarak2017}
			Rivkind A, Barak O. 2017 
			Local dynamics in trained recurrent neural networks. 
			Phys. Rev. Let., 118(25), p.258101.

		\bibitem{CeniLivi2018}
			Ceni A, Ashwin P, Livi L. 2018 
			Interpreting RNN behaviour via excitable network attractors. 
			arXiv preprint arXiv:1807.10478.

		\bibitem{MaassBertschinger2005}
			Maass W, Legenstein RA, Bertschinger N. 2005 
			Methods for estimating the computational power and generalization capability of neural microcircuits. 
			In Advances in neural information processing systems, 865-872.

		\bibitem{LegensteinMaass2007b}
			Legenstein R, Maass W. 2007 
			Edge of chaos and prediction of computational performance for neural circuit models. 
			Neural Networks 20(3), 323-334.

		\bibitem{HauserMaass2011}
			Hauser H, Ijspeert AJ, F\"uchslin RM, Pfeifer R, Maass W. 2011 
			Towards a theoretical foundation for morphological computation with compliant bodies. 
			Biol. Cybern. 105, 355-370. 
		
		\bibitem{NakajimaPfeifer2013a}
			Nakajima K, Hauser H, Kang R, Guglielmino E, Caldwell DG, Pfeifer R. 2013 
			A soft body as a reservoir: case studies in a dynamic model of octopus-inspired soft robotic arm. 
			Front. Comput. Neurosc. 7, 91.

		\bibitem{NicheleGundersen2017}
			Nichele S, Gundersen MS. 2017 
			Reservoir Computing Using Non-Uniform Binary Cellular Automata. 
			arXiv preprint arXiv:1702.03812.

		\bibitem{SorianoVanDerSande2015}
			Soriano MC, Ort\'in S, Keuninckx L, Appeltant L, Danckaert J, Pesquera L, Van der Sande G. 2015 
			Delay-based reservoir computing: noise effects in a combined analog and digital implementation. 
			IEEE transactions on neural networks and learning systems, 26(2), pp.388-393.

		\bibitem{DuLu2017}
			Du C, Cai F, Zidan MA, Ma W, Lee SH, Lu WD. 2017 
			Reservoir computing using dynamic memristors for temporal information processing. 
			Nat. Com. 8(1), p.2204.

		\bibitem{FernandoSojakka2003}
			Fernando C, Sojakka S. 2003 
			Pattern recognition in a bucket. 
			In European conference on artificial life (pp. 588-597). Springer, Berlin, Heidelberg.

		\bibitem{AppletantFischer2011}
			Appeltant L, Soriano MC, Van der Sande G, Danckaert J, Massar S, Dambre J, Schrauwen B, Mirasso CR, Fischer I. 2011 
			Information processing using a single dynamical node as complex system. 
			Nat. Commun. 2, 468.

		\bibitem{PaquotMassar2012}
			Paquot Y, Duport F, Smerieri A, Dambre J, Schrauwen B, Haelterman M, Massar S. 2012 
			Optoelectronic reservoir computing. 
			Sci. Rep. 2, 287.

		\bibitem{VandoorneBienstman2014}
			Vandoorne K, Mechet P, Van Vaerenbergh T, Fiers M, Morthier G, Verstraeten D, Schrauwen B, Dambre J, Bienstman P. 2014 
			Experimental demonstration of reservoir computing on a silicon photonics chip. 
			Nat. Commun., 5, p.3541.

		\bibitem{LegensteinMaass2007a}
			Legenstein R, Maass W. 2007 
			What makes a dynamical system computationally powerful. 
			New directions in statistical signal processing: From systems to brain, 127-154.

		\bibitem{Vapnik1998}
			Vapnik V. 1998 
			Statistical learning theory. 
			Wiley, New York.

		\bibitem{CherkasskyMulier1998}
			Cherkassky V, Mulier F. 1998 
			Learning from data: Concepts, theory, and methods. 
			New York: Wiley.

		\bibitem{MarkramTsodyks1998}
			Markram H, Wang Y, Tsodyks M. 1998 
			Differential signaling via the same axon of neocortical pyramidal neurons. 
			Proc. Nat. Acad. Sci. 95(9), 5323-5328.

		\bibitem{Coello2006}
			Coello CC. 2006 
			Evolutionary multi-objective optimization: a historical view of the field. 
			IEEE computational intelligence magazine, 1(1), pp.28-36.

		\bibitem{Schuster2012}
			Schuster P. 2012 
			Optimization of multiple criteria: Pareto efficiency and fast heuristics should be more popular than they are. 
			Complexity, 18(2), pp.5-7.

		\bibitem{Seoane2016}
			Seoane LF. 2016 
			Multiobjetive optimization in models of synthetic and natural living systems. 
			PhD Thesis, Universitat Pompeu Fabra.

		\bibitem{ShovalAlon2012}
			Shoval O, Sheftel H, Shinar G, Hart Y, Ramote O, Mayo A, Dekel E, Kavanagh K, Alon U. 2012 
			Evolutionary trade-offs, Pareto optimality, and the geometry of phenotype space. 
			Science, p.1217405.

		\bibitem{HartAlon2015}
			Hart Y, Sheftel H, Hausser J, Szekely P, Ben-Moshe NB, Korem Y, Tendler A, Mayo AE, Alon U. 2015 
			Inferring biological tasks using Pareto analysis of high-dimensional data. 
			Nat. Methods, 12(3), p.233.

		\bibitem{SzekelyAlon2015}
			Szekely P, Korem Y, Moran U, Mayo A, Alon U. 2015 
			The mass-longevity triangle: Pareto optimality and the geometry of life-history trait space. 
			PLoS Comp. Biol., 11(10), p.e1004524.

		\bibitem{TendlerAlon2015}
			Tendler A, Mayo A, Alon U. 2015 
			Evolutionary tradeoffs, Pareto optimality and the morphology of ammonite shells. 
			BMC Syst. Biol., 9(1), p.12.

		\bibitem{SeoaneSole2013}
			Seoane LF, Sol\'e R. 2013 
			A multiobjective optimization approach to statistical mechanics. 
			arXiv preprint arXiv:1310.6372.

		\bibitem{SeoaneSole2015a}
			Seoane LF, Sol\'e' R. 2015 
			Phase transitions in Pareto optimal complex networks. 
			Physical Review E, 92(3), p.032807.

		\bibitem{SeoaneSole2016}
			Seoane LF, Sol\'e R. 2016 
			Multiobjective optimization and phase transitions. 
			In Proceedings of ECCS 2014 (pp. 259-270). Springer, Cham.

		\bibitem{SeoaneSole2015b}
			Seoane LF, Sol\'e' R. 2015 
			Systems poised to criticality through Pareto selective forces. 
			arXiv preprint arXiv:1510.08697.

		\bibitem{Bak1996}
			Bak P. 1996 
			How nature works: the science of self-organized criticality. 
			Springer Science \& Business Media.

		\bibitem{BeggPlenz2003}
			Beggs JM, Plenz D. 2003 
			Neuronal avalanches in neocortical circuits. 
			J. Neurosci., 23(35), pp.11167-11177.

		\bibitem{Chialvo2010}
			Chialvo DR. 2010 
			Emergent complex neural dynamics. 
			Nat. Phys., 6(10), p.744.

		\bibitem{MoraBialek2011}
			Mora T, Bialek W. 2011 
			Are biological systems poised at criticality?. 
			J. Stat. Phys., 144(2), pp.268-302.

		\bibitem{TagliazucchiChialvo2012}
			Tagliazucchi E, Balenzuela P, Fraiman D, Chialvo DR. 2012 
			Criticality in large-scale brain fMRI dynamics unveiled by a novel point process analysis. 
			Front. Physiol., 3, p.15.

		\bibitem{MorettiMunoz2013}
			Moretti P, Mu\~noz MA. 2013 
			Griffiths phases and the stretching of criticality in brain networks. 
			Nat. Com., 4, 2521.

		\bibitem{Munoz2018}
			Munoz MA. 2018  
			{\em Colloquium}: Criticality and dynamical scaling in living systems. 
			Rev. Mod. Phys., 90(3), 031001.

		\bibitem{Wolfram1984}
			Wolfram S. 1984 
			Universality and complexity in cellular automata. 
			Physica D, 10(1-2), 1-35.

		\bibitem{Langton1990}
			Langton CG. 1990 
			Computation at the edge of chaos: phase transitions and emergent computation. 
			Physica D, 42(1-3), 12-37.

		\bibitem{MitchellCrutchfield1993}
			Mitchell M, Hraber P, Crutchfield JP. 1993 
			Revisiting the edge of chaos: Evolving cellular automata to perform computations. 
			arXiv preprint adap-org/9303003.

		\bibitem{BertschingerNatschlager2004}
			Bertschinger N, Natschl\"ager T. 2004 
			Real-time computation at the edge of chaos in recurrent neural networks. 
			Neural Computation, 16(7), 1413-1436.

		\bibitem{SchrauwenLegenstein2009}
			Schrauwen B, Büsing L, Legenstein RA. 2009 
			On computational power and the order-chaos phase transition in reservoir computing. 
			In Advances in Neural Information Processing Systems (pp. 1425-1432).

		\bibitem{ToyozumiAbbott2011}
			Toyoizumi T, Abbott LF. 2011 
			Beyond the edge of chaos: Amplification and temporal integration by recurrent networks in the chaotic regime. 
			Physical Review E, 84(5), 051908.

		\bibitem{BoedeckerAsada2012}
			Boedecker J, Obst O, Lizier JT, Mayer NM, Asada M. 2012 
			Information processing in echo state networks at the edge of chaos. 
			Theory in Biosciences, 131(3), pp.205-213.

		\bibitem{BianchiAlippi2018a}
			Bianchi FM, Livi L, Alippi C. 2018 
			Investigating echo-state networks dynamics by means of recurrence analysis. 
			IEEE transactions on neural networks and learning systems, 29(2), pp.427-439.

		\bibitem{BianchiAlippi2018b}
			Bianchi FM, Livi L, Alippi C. 2018 
			On the Interpretation and Characterization of Echo State Networks Dynamics: A Complex Systems Perspective. 
			In Advances in Data Analysis with Computational Intelligence Methods (pp. 143-167). Springer, Cham.

		\bibitem{LiviAlippi2018}
			Livi L, Bianchi FM, Alippi C. 2018 
			Determination of the edge of criticality in echo state networks through Fisher information maximization. 
			IEEE Transactions on Neural Networks and Learning Systems, 29(3), pp.706-717.

		\bibitem{ProkopenkoWang2011}
			Prokopenko M, Lizier JT, Obst O, Wang XR. 2011 
			Relating Fisher information to order parameters. 
			Phys. Rev. E, 84(4), 041116.

		\bibitem{MaassMarkram2004a}
			Maass W, Natschl\"ager T and Markram H. 2004 
			Computational models for generic cortical microcircuits. 
			Comp. Neurosci., 18, p.575.

		\bibitem{MaassMarkram2004b}
			Maass W, Natschl\"ager T, Markram H. 2004 
			Fading memory and kernel properties of generic cortical microcircuit models. 
			J. Physiol.-Paris, 98(4-6), pp.315-330.

		\bibitem{MaassMarkram2006}
			Maass W, Markram H. 2006 
			Theory of the computational function of microcircuit dynamics. 
			In The interface between neurons and global brain function, Dahlem Workshop Report (Vol. 93, pp. 371-390).

		\bibitem{ThomsonBannister2002}
			Thomson AM, West DC, Wang Y, Bannister AP. 2002 
			Synaptic connections and small circuits involving excitatory and inhibitory neurons in layers 2-5 of adult rat and cat neocortex: triple intracellular recordings and biocytin labelling in vitro. 
			Cerebral cortex, 12(9), pp.936-953.

		\bibitem{HubelWiesel1962}
			Hubel DH, Wiesel TN. 1962 
			Receptive fields, binocular interaction and functional architecture in the cat's visual cortex. 
			J. Physiol., 160(1), 106-154.

		\bibitem{DiamondAhissar2008}
			Diamond ME, Von Heimendahl M, Knutsen PM, Kleinfeld D, Ahissar E. 2008 
			`Where' and `what' in the whisker sensorimotor system. 
			Nat. Rev. Neurosci., 9(8), 601.

		\bibitem{HawkinsBlakeslee2007}
			Hawkins J, Blakeslee S. 2007 
			On intelligence: How a new understanding of the brain will lead to the creation of truly intelligent machines. 
			Macmillan.

		\bibitem{OberlaenderSakmann2011}
			Oberlaender M, de Kock CP, Bruno RM, Ramirez A, Meyer HS, Dercksen VJ, Helmstaedter M, Sakmann B. 2011 
			Cell type-specific three-dimensional structure of thalamocortical circuits in a column of rat vibrissal cortex. 
			Cereb. Cortex 22(10), pp.2375-2391.

		\bibitem{HabenschussMaass2013}
			Habenschuss S, Jonke Z,  Maass W. 2013 
			Stochastic computations in cortical microcircuit models. 
			PLoS Comp Biol 9(11), p.e1003311.

		\bibitem{HaeuslerMaass2006}
			Haeusler S, Maass W. 2006 
			A statistical analysis of information-processing properties of lamina-specific cortical microcircuit models. 
			Cereb. Cortex 17(1), pp.149-162.

		\bibitem{Maass2016}
			Maass W. 2016 
			Searching for principles of brain computation. 
			Current Opinion in Behavioral Sciences, 11, 81-92.

		\bibitem{BernacchiaWang2011}
			Bernacchia A, Seo H, Lee D, Wang XJ. 2011 
			A reservoir of time constants for memory traces in cortical neurons. 
			Nat. Neurosci., 14(3), 366.

		\bibitem{NikolicMaass2009}
			Nikoli\'c D, H\"ausler S, Singer W, Maass W. 2009 
			Distributed fading memory for stimulus properties in the primary visual cortex. 
			PLoS Biol., 7(12), p.e1000260.

		\bibitem{DraniasVanDongen2013}
			Dranias MR, Ju H, Rajaram E, VanDongen AM. 2013 
			Short-term memory in networks of dissociated cortical neurons. 
			J. Neurosci., 33(5), pp.1940-1953.

		\bibitem{JuVanDongen2015}
			Ju H, Dranias MR, Banumurthy G, VanDongen, AM. 2015 
			Spatiotemporal memory is an intrinsic property of networks of dissociated cortical neurons. 
			J. Neurosci., 35(9), 4040-4051.

		\bibitem{MarreBerry2015}
			Marre O, Botella-Soler V, Simmons KD, Mora T, Tka\v{c}ik G, Berry II MJ. 2015 
			High accuracy decoding of dynamical motion from a large retinal population. 
			PLoS Comp. Biol., 11(7), p.e1004304.

		\bibitem{Singer2013}
			Singer W. 2013 
			Cortical dynamics revisited. 
			Trends in cognitive sciences, 17(12), 616-626.

		\bibitem{KlampflMaass2013}
			Klampfl S, Maass W. 2013 
			Emergence of dynamic memory traces in cortical microcircuit models through STDP. 
			J. Neurosci., 33(28), 11515-11529.

		\bibitem{ChenHelmchen2013}
			Chen JL, Carta S, Soldado-Magraner J, Schneider BL, Helmchen F. 2013 
			Behaviour-dependent recruitment of long-range projection neurons in somatosensory cortex. 
			Nature, 499(7458), 336.

		\bibitem{KlampflMaass2012}
			Klampfl S, David SV, Yin P, Shamma SA, Maass W. 2012 
			A quantitative analysis of information about past and present stimuli encoded by spikes of A1 neurons. 
			J. Neurophysiol., 108(5), 1366-1380.

		\bibitem{RigottiFusi2013}
			Rigotti M, Barak O, Warden MR, Wang XJ, Daw ND, Miller EK, Fusi S. 2013 
			The importance of mixed selectivity in complex cognitive tasks. 
			Nature, 497(7451), 585.

		\bibitem{FusiRigotti2016}
			Fusi S, Miller EK, Rigotti M. 2016 
			Why neurons mix: high dimensionality for higher cognition. 
			Current opinion in neurobiology, 37, 66-74.

		\bibitem{NakajimaPfeifer2013b}
			Nakajima K, Hauser H, Kang R, Guglielmino E, Caldwell DG, Pfeifer R. 2013 
			Computing with a muscular-hydrostat system. 
			In Robotics and Automation (ICRA), 2013 IEEE International Conference on (pp. 1504-1511). IEEE.

		\bibitem{GodfreySmith2016}
			Godfrey-Smith P. 2016 
			Other minds: The octopus, the sea, and the deep origins of consciousness. 
			Farrar, Straus and Giroux.

		\bibitem{Sacks2017}
			Sacks O. 2017 
			The river of consciousness, Ch. 3. 
			Picador.

		\bibitem{PfeiferBongard2006}
			Pfeifer R, Bongard J. 2006 
			How the body shapes the way we think: a new view of intelligence. 
			MIT press.

		\bibitem{PfeiferIida2007}
			Pfeifer R, Lungarella M, Iida F. 2007 
			Self-organization, embodiment, and biologically inspired robotics. 
			Science 318(5853), 1088-1093.

		\bibitem{MullerHoffman2017}
			Müller VC, Hoffmann M. 2017 
			What is morphological computation? On how the body contributes to cognition and control. 
			Artif. Life 23(1), 1-24.

		\bibitem{McGeer1990}
			McGeer T. 1990 
			Passive dynamic walking. 
			I. J. Robotic Res. 9(2), 62-82.

		\bibitem{WisseVanFrankenhuyzen2006}
			Wisse M, Van Frankenhuyzen J. 2006 
			Design and construction of MIKE; a 2-D autonomous biped based on passive dynamic walking. 
			In Adaptive motion of animals and machines (pp. 143-154). 
			Springer, Tokyo.

		\bibitem{FendPfeifer2006}
			Fend M, Bovet S, Pfeifer R. 2006 
			On the influence of morphology of tactile sensors for behavior and control. 
			Robot. Auton. Syst., 54(8), 686-695.

		\bibitem{MurataKurokawa2007}
			Murata S, Kurokawa H. 2007 
			Self-reconfigurable robots. 
			IEEE Robot. Autom. Mag., 14(1), 71-78.

		\bibitem{HauserMaass2012}
			Hauser H, Ijspeert AJ, F\"uchslin RM, Pfeifer R, Maass W. 2012 
			The role of feedback in morphological computation with compliant bodies. 
			Biol. Cybern. 106(10), 595-613.

		\bibitem{SumiokaPfeifer2011}
			Sumioka H, Hauser H, Pfeifer R. 2011 
			Computation with mechanically coupled springs for compliant robots. 
			In Intelligent Robots and Systems (IROS), 2011 IEEE/RSJ International Conference on (pp. 4168-4173). IEEE.

		\bibitem{NakajimaPfeifer2013c}
			Nakajima K, Hauser H, Kang R, Guglielmino E, Caldwell DG, Pfeifer R. 2013 
			Computing with a muscular-hydrostat system. 
			In Robotics and Automation (ICRA), 2013 IEEE International Conference on (pp. 1504-1511). IEEE.

		\bibitem{CaluwaertsSchrauwen2011}
			Caluwaerts K, Schrauwen B. 2011 
			The body as a reservoir: locomotion and sensing with linear feedback. 
			In 2nd International conference on Morphological Computation (ICMC 2011).

		\bibitem{CaluwaertsSchrauwen2013}
			Caluwaerts K, D'Haene M, Verstraeten D, Schrauwen B. 2013 
			Locomotion without a brain: physical reservoir computing in tensegrity structures. 
			Artif. Life 19(1), 35-66.

		\bibitem{GabaldaSagarraGarciaOjalvo2018}
			Gabalda-Sagarra M, Carey LB, Garc\'ia-Ojalvo J. 2018 
			Recurrence-based information processing in gene regulatory networks. 
			Chaos 28(10), p.106313.

		\bibitem{Clark1998}
			Clark A. 1998 
			Being there: Putting brain, body, and world together again. 
			MIT press.

		\bibitem{NakajimaPfeifer2014}
			Nakajima K, Li T, Hauser H, Pfeifer R. 2014 
			Exploiting short-term memory in soft body dynamics as a computational resource. 
			J. R. Soc. Interface, 11(100), 20140437.

		\bibitem{NakajimaPfeifer2015}
			Nakajima K, Hauser H, Li T, Pfeifer R. 2015 
			Information processing via physical soft body. 
			Sci. Rep., 5, 10487. 

		\bibitem{ShimHusbands2007}
			Shim Y, Husbands P. 2007 
			Feathered flyer: integrating morphological computation and sensory reflexes into a physically simulated flapping-wing robot for robust flight manoeuvre. 
			In European Conference on Artificial Life (pp. 756-765). Springer, Berlin, Heidelberg.

		\bibitem{ValeroCuevasLipson2007}
			Valero-Cuevas FJ, Yi JW, Brown D, McNamara RV, Paul C, Lipson H. 2007 
			The tendon network of the fingers performs anatomical computation at a macroscopic scale. 
			IEEE T. Bio-Med. Eng. 54(6), 1161-1166.

		\bibitem{Minsky2017}
			Minsky M, Papert SA. 2017 
			Perceptrons: An introduction to computational geometry. 
			MIT press.

		\bibitem{Hopfield1982}
			Hopfield JJ. 1982 
			Neural networks and physical systems with emergent collective computational abilities. 
			Proc. Nat. Acad. Sci. 79(8), 2554-2558. 

		\bibitem{Kohonen1982}
			Kohonen T. 1982 
			Self-organized formation of topologically correct feature maps. 
			Biol. Cybern. 43(1), 59-69.

		\bibitem{Hebb1963}
			Hebb DO. 1963 
			The Organizations of Behavior: a Neuropsychological Theory. 
			Lawrence Erlbaum.

		\bibitem{Ritter1990}
			Ritter H. 1990 
			Self-organizing maps for internal representations. 
			Psychol. Res., 52(2-3), 128-136.

		\bibitem{SpitzerKischka1995}
			Spitzer M, Bh\"oler P, Weisbrod M, Kischka U. 1995 
			A neural network model of phantom limbs. 
			Biol. Cybern., 72(3), pp.197-206.

		\bibitem{Raup1966}
			Raup DM. 1966 
			Geometric analysis of shell coiling: general problems. 
			Journal of Paleontology, pp.1178-1190.

		\bibitem{Niklas1997}
			Niklas KJ. 1997 
			The evolutionary biology of plants. 
			Chicago U. Press. 

		\bibitem{McGhee1999}
			McGhee GR. 1999 
			Theoretical morphology: the concept and its applications. 
			Columbia U. Press. 

		\bibitem{Niklas2004}
			Niklas KJ. 2004 
			Computer models of early land plant evolution. 
			Annu. Rev. Earth Planet. Sci., 32, 47-66.

		\bibitem{CorominasMurtraRodriguezCaso2013}
			Corominas-Murtra B, Go\~ni J, Sol\'e RV and Rodr\'iguez-Caso C. 2013 
			On the origins of hierarchy in complex networks. 
			Proc. Nat. Acad. Sci., 110(33), 13316-13321.

		\bibitem{AvenaKoenigsbergerSporns2015}
			Avena-Koenigsberger A, Go\~ni J, Sol\'e R and Sporns O. 2015 
			Network morphospace. 
			J. R. Soc. Interface, 12(103), 20140881.

		\bibitem{SeoaneSole2018b}
			Seoane LF, Sol\'e R. 2018 
			The morphospace of language networks. 
			Sci. Rep., 8, 10465. 

		\bibitem{YusteLansner2005}
			Yuste R, MacLean JN, Smith J, Lansner A. 2005 
			The cortex as a central pattern generator. 
			Nat. Rev. Neurosci. 6(6), 477-483.

		\bibitem{MarderCalabrese1996}
			Marder E, Calabrese RL. 1996 
			Principles of rhythmic motor pattern generation. 
			Physiol. Rev., 76(3), 687-717.

		\bibitem{Ijspeert2008}
			Ijspeert AJ. 2008 
			Central pattern generators for locomotion control in animals and robots: a review. 
			Neural Networks, 21(4), 642-653.

		\bibitem{BaluskaLevin2016}
			Balu\v{s}ka F, Levin M. 2016 
			On having no head: cognition throughout biological systems. 
			Front. Psychol., 7, p.902.


	\end{thebibliography}
\end{document}